%% file: arxiv.tex
\documentclass[sigconf]{aamas} 

\input{preamble}

\begin{document}


\pagestyle{fancy}
\fancyhead{}

\maketitle 

\input{main_body}

\include{appendix}

\end{document}

%% file: preamble.tex
\usepackage{balance} 
\usepackage{graphicx}
\usepackage{booktabs} 
\usepackage[ruled]{algorithm2e} 
\usepackage{xcolor} 
\usepackage{soul}
\usepackage{amsmath, amsfonts}
\usepackage{tabularx}
\usepackage{natbib}
\usepackage{multirow}
\usepackage{hyperref}
\usepackage{xr-hyper}

\doi{BAHE1423}

\makeatletter
\gdef\@copyrightpermission{
  \begin{minipage}{0.2\columnwidth}
   \href{https://creativecommons.org/licenses/by/4.0/}{\includegraphics[width=0.90\textwidth]{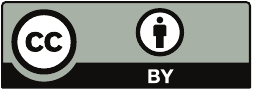}}
  \end{minipage}\hfill
  \begin{minipage}{0.8\columnwidth}
   \href{https://creativecommons.org/licenses/by/4.0/}{This work is licensed under a Creative Commons Attribution International 4.0 License.}
  \end{minipage}
  \vspace{5pt}
}
\makeatother

\setcopyright{ifaamas}
\acmConference[AAMAS '26]{Proc.\@ of the 25th International Conference
on Autonomous Agents and Multiagent Systems (AAMAS 2026)}{May 25 -- 29, 2026}
{Paphos, Cyprus}{C.~Amato, L.~Dennis, V.~Mascardi, J.~Thangarajah (eds.)}
\copyrightyear{2026}
\acmYear{2026}
\acmDOI{}
\acmPrice{}
\acmISBN{}

\acmSubmissionID{1373}

\title{Learning Bayesian Game Families, with Application to Mechanism Design}

\author{Madelyn Gatchel}
\affiliation{
  \institution{University of Michigan}
  \city{Ann Arbor, MI}
  \country{USA}}
\email{gatchel@umich.edu}

\author{Michael P. Wellman}
\affiliation{
  \institution{University of Michigan}
  \city{Ann Arbor, MI}
  \country{USA}}
\email{wellman@umich.edu}

\begin{abstract}
Learning or estimating game models from data typically entails inducing separate models for each setting, even if the games are parametrically related.
In empirical mechanism design, for example, this approach requires learning a new game model for each candidate setting of the mechanism parameter. 
Recent work has shown the data efficiency benefits of learning a single parameterized model for families of related games. 
In \textit{Bayesian games}---a typical model for mechanism design---payoffs depend on both the actions and types of the players.
We show how to exploit this structure by learning an \textit{interim} game-family model that conditions on a single player’s type. 
We compare this to the baseline approach of directly learning the \textit{ex ante} payoff function, which gives payoffs in expectation of all player types. 
By marginalizing over player type, the interim model can also provide ex ante payoff predictions, as necessary for Bayes-Nash equilibrium approximation. 
We also leverage the interim model to compute new beneficial piecewise best-response strategies, without any additional sample data.
We validate our method through a case study of a dynamic sponsored search auction. For both payoff accuracy and Nash-approximation error, the interim model matches the ex ante model on the trained range, and outperforms ex ante in extrapolation.
Our case study demonstrates that Bayesian game-family models can support comprehensive mechanism design, and that through interim-stage modeling we can enhance expressivity and reliability.  
\end{abstract}


\newcommand{\BibTeX}{\rm B\kern-.05em{\sc i\kern-.025em b}\kern-.08em\TeX}
\newcommand{\abs}[1]{\left| #1 \right|}
\newcommand{\dotproduct}{\kern-.06em \boldsymbol{\cdot} \kern-.06em}

\newcommand{\idx}[2]{#1^{(#2)}}

\newcommand{\vidx}[2]{\vec{#1}^{(#2)}}

\newcommand{\vecsp}[1]{\vec{#1}\mkern2mu}

\newcommand{\term}[1]{\textbf{\textit{#1}}}
\newcommand{\nmix}{m}
\newcommand{\nobs}{o}
\newcommand{\nsimq}{Q}
\newcommand{\val}{\theta}
\newcommand{\reserve}{r}

\newcommand{\partition}{\mathcal{C}}
\newcommand{\interval}{C}
\newcommand{\typesamp}{\mathcal{N}}
\newcommand{\BNE}{\mathit{BNE}}
\newcommand{\pbr}{\psi}

\newcommand{\maxr}{\bar{r}}
\DeclareMathOperator*{\argmax}{\arg\,max}
\providecommand{\AAMASARXIVNOTE}{}
\providecommand{\LOADAPPENDIXREFS}{} 

\LOADAPPENDIXREFS

%% file: main_body.tex


\section{Introduction}

Many real-world strategic interactions can be modeled as \term{Bayesian games}, where payoffs depend on players’ actions and their private information, or \term{types}. 
Outcomes may also depend on environment parameters, such that each parameter setting induces a different Bayesian game. 
Given limited modeling resources, analysts must decide in advance the parameter range and its granularity.  
While domain knowledge may guide this selection, it cannot guarantee coverage of all salient settings. 
Ideally, analysts would have a model of the \term{Bayesian game family} from which they could reason about any relevant parameter setting.


A motivating application for Bayesian game families is \term{mechanism design}, where a 
designer sets or influences an environment parameter that affects strategic incentives. 
Each parameter value results in a different \term{game instance}.
The designer's goal is to find the parameter setting that optimizes a relevant objective function, such as social welfare or revenue. 
In \term{empirical mechanism design} (EMD), game model instances are induced from simulation data.
Past EMD studies \citep{brinkman2017empirical,vorobeychik2006empirical, jordan2010strategy} have generally focused on a limited set of mechanism settings, separately modeling and analyzing each game instance. 

\citet{gatchel2023learning} demonstrated that learning a single parameterized payoff model for families of related normal-form games is more data-efficient than training separate models for each game instance. 
We extend this approach to Bayesian game families, exploiting the type-conditional form of strategies for these games.
Specifically, we investigate the learning of \term{interim payoff functions}, which explicitly condition on a single player's type.
By marginalizing out this type, we obtain the \term{ex ante payoff functions} which are essentially the payoffs learned in a normal-form model. 
We also explore learning ex ante payoff functions directly, and compare ex ante and interim learning approaches.

We validate our method through an EMD case study in the domain of sponsored search \citep{lahaie2007sponsored}, where the publisher sets an auction reserve requirement to maximize revenue in equilibrium.
Our search auction model is designed to capture the dynamic nature of bidding, where advertisers can revise their bids based on provisional results of earlier bidding rounds. 
We do so in a two-stage scenario, in which the bidders can \emph{attempt} to modify their bids given the state of bidding after the first round.
These attempts succeed probabilistically, thus providing an incentive for the players to submit meaningful first-round bids.
The scenario is simple to describe and design heuristic strategies, yet too complex for straightforward analytic solution.
We implement an agent-based simulation model of the scenario, and from the simulation-generated data learn Bayesian game-family models to support empirical game-theoretic analysis and mechanism design.

In our experiments, both ex ante and interim models achieve low payoff error on the trained parameter range, yet only the interim model maintains low error in extrapolation. 
Equilibria approximated using the interim model have regret error comparable to ex ante within the trained range and lower error beyond it.
We also demonstrate that the learned game-family models support effective EMD procedures.
Analysis of the expected revenue curve using a fine-grained parameter grid reveals the benefit of learning multiple game-family models.
This analysis also provides concrete evidence that the interim model's extrapolation capability is advantageous for mechanism design. 
A final advantage of the interim model is its ability to generate new strategies that outperform those in the original set. 
We introduce \textit{piecewise-conditional} strategies that select from the original strategies based on the bidder's own type, and we show how to construct such strategies that beneficially respond to equilibria over the original strategy set.
Our investigation produces new insights about alternative model forms for Bayesian game families, and provides compelling experimental evidence in favor of learning interim payoff functions. 

\section{Related Work}

\subsection{Learning Game Models from Data}

In \term{empirical game-theoretic analysis} (EGTA) \citep{wellman2025empirical}, complex multi-agent interactions are represented by an agent-based simulation model.
Data from agent-based simulation is used to used to induce a formal game model, called an \term{empirical game}, which can be analyzed using standard methods, for example to derive approximate equilibria.%
\footnote{Games defined in terms of simulation models have also been called \term{simulation-based} \citep{areyanviqueira2020improved,marchesi2020learning,sokota2019learning} or \term{black-box games} \citep{zhang2021finding}.}
Given limited sampling resources, it is infeasible to estimate payoffs for every strategy profile from simulation data.

An alternative is to learn game models that generalize from the available data, such as using regression to learn pure-strategy payoff functions  \citep{vorobeychik2007learning, wiedenbeck2018regression} or inducing compact game structures \citep{ficici2008learning, duong2009learning, li2020structure, liu2023nfgtransformer}.
From a theoretical perspective, \citet{fu2020learning} demonstrate that in non-truthful auctions, the interim utility function admits polynomial sample complexity, but the ex ante utility function does not. 
\citet{sokota2019learning} train a neural network to learn the \term{deviation payoff function}---the expected utility for a unilateral deviator---for symmetric normal-form empirical game instances.
This technique leverages the compactness of payoff functions in role-symmetric games. 
The learned deviation payoff function supports equilibrium computation without the combinatorial mixture summations required for direct payoff functions.
\citet{li2021evolution} learn deviation payoffs as part of a method to approximate mixed-strategy Bayes-Nash equilibria through iterative evolutionary search. 
\citet{gatchel2023learning} learn the deviation payoff function for families of related symmetric normal-form game instances, showing that this approach achieves higher payoff accuracy with less data than learning separate models for each game instance.

\subsection{Empirical Mechanism Design}

\citet{vorobeychik2006empirical} analyze five game instances with different storage costs in a supply-chain management scenario and find that none deter initial procurement without sacrificing profitability. 
\citet{jordan2010strategy} examine how reserve scores impact revenue in online advertising auctions, analyzing 14 game instances and identifying the revenue-maximizing score. 
\citet{brinkman2017empirical} investigate the optimal clearing interval in a frequent call market to maximize allocative efficiency as the number of agents and trade opportunities are varied. 
Most recently, \citet{wang2025empirical} evaluate the effectiveness of different security levels at thwarting cyber attacks in mixnets, analyzing four game instances as part of an empirical mechanism design study. 


\citet{vorobeychik2012constrained} present an automated mechanism design procedure, framing EMD as a black-box optimization problem compatible with any evaluation method.
Evaluating a candidate mechanism setting is a subproblem that involves approximating an equilibrium in the induced game and computing the objective in equilibrium. 
\citet{areyanviqueira2020empirical} introduce a PAC-learning EMD framework that includes approximate equilibrium estimation and parameter search with Bayesian optimization.
\citet{zhang2023computing} reformulate multi-player extensive-form mechanism design problems as 2-player zero-sum games and solve for optimal equilibria via learning. 

Other approaches have used neural networks to learn mechanism solutions without explicit game modeling. 
\citet{dutting2019optimal} train allocation and payment neural networks to learn auction rules that are incentive compatible. 
\citet{bichler2021learning} employ self-play policy iteration with neural networks to learn approximate Bayes-Nash equilibrium bidding strategies in symmetric auction games. 
\citet{gemp2022designing} learn the auctioneer's utility as a function of contest design in empirical all-pay auctions or crowdsourcing contests. 



\section{Background}


A \term{symmetric Bayesian game} $\Gamma$ has $p$ players who share the same action set $A$, type space $T$, and utility function $U$. 
In this work, we assume that $A$ is finite, and that $T$ is infinite and ordered.
Each player has a private type $t \in T$, which we assume is drawn independently from a common probability distribution  $\mu\in\Delta(T)$.%
\footnote{Note that the player symmetry is \term{ex ante} (prior to drawing types); once types are drawn the players are in distinct situations.}
The utility function is given by $U: A^p\times T^p \to \mathbb{R}$, where the payoff for a player playing action $a$ with type $t$ while opponents with types $\vec{t}$ play actions $\vec{a}$, is denoted as $U\big((a, \vec{a}),  (t, \vecsp{t})\big)$. 
Vectors $\vec{t}$ and $\vec{a}$ have length $p-1$, have a consistent order,  and specify the corresponding types and actions of the $p-1$ opponents.
To improve readability, we omit the inner parentheses and write $U(a, \vec{a}, t, \vecsp{t})$. 

Let $S$ be the set of pure strategies, each strategy $s\in S$ a function $s: T \to A$.
We assume $S$ is finite, thus a strict (typically highly restricted) subset of all mappings from types to actions.
An \term{opponent profile} is a length-$(p-1)$ vector $\vec{s} \in \vec{S}$ that specifies each opponent's strategy, where the order of entries matches the order of opponents in the type vector $\vec{t}$.
We assume access to an oracle (e.g., a simulator) that gives a noisy payoff estimate for a player with type $t$ who plays strategy~$s_j$, given $(\vec{s}, \vecsp{t})$: 
$\tilde{U}\big(s_j(t), \vec{s}(\vecsp{t}), t, \vecsp{t}\big).$
Given player symmetry, we can define a utility function $u_j$ \textit{for each strategy} $s_j$ that maps opponent profiles to payoffs.
We first define the \term{ex ante expected payoff}: 
\begin{equation}u_j(\vec{s}) = \mathbb{E}_{(t, \vec{t})\sim \mu^p} \Big[U\big(s_j(t), \vec{s}(\vecsp{t}), \; t, \vecsp{t} \big) \Big].
\label{eq:ex_ante_exp_util}
\end{equation} 
In the \term{interim} stage, each player knows its own type but has only probabilistic beliefs about the others.
The \term{interim expected payoff} is given by
\begin{equation}u_j(\vec{s} \mid t) = \mathbb{E}_{\vec{t}\sim \mu^{p-1}} \Big[U\big(s_j(t), \vec{s}(\vecsp{t}), t, \vecsp{t} \big) \Big].
\label{eq:ex_interim_exp_util}
\end{equation} 
We can express ex ante as a marginalized version of interim: 
$u_j(\vec{s}) = \mathbb{E}_{t}u_j(\vec{s} \mid t)$.

We learn the deviation payoff function, a mapping from mixed-strategy profiles to vectors of expected utilities for unilateral strategy deviations.
Let $\sigma \in \Delta(S)$ denote a mixed strategy, where $\Delta(S)$ is the probability simplex over the strategy set $S$.
Further, let $\vec{\sigma}$ refer to a symmetric mixed-strategy profile where all $p$ (or $p-1$) players are playing according to~$\sigma$.
We define the \term{ex ante} and \term{interim deviation payoffs} for deviating to strategy $s_j$: 
\begin{align} u_j(\vec{\sigma}) &= \sum_{\vec{s} \in \vec{S}} \Pr(\vec{s} \mid \vec{\sigma}) u_j(\vec{s}), \label{eq:devpay_exante}\\
u_j(\vec{\sigma}\mid t) &= \sum_{\vec{s} \in \vec{S}} \Pr(\vec{s} \mid \vec{\sigma}) u_j(\vec{s} \mid t). \label{eq:devpay_exinterim}
\end{align}
We can write the ex ante deviation payoff for strategy $s_j$ using the interim deviation payoff for strategy $s_j$:
\[u_j(\vec{\sigma}) = \mathbb{E}_{t\sim \mu} u_j(\vec{\sigma} \mid t) = \int_t u_j(\vec{\sigma}\mid t)\mu(t)dt.\]

The \term{ex ante deviation payoff function} is given by
the $\abs{S}$-dimensional vector $u(\vec{\sigma}) = [u_j(\vec{\sigma})]$, over $s_j \in S$.
Similarly, the \term{interim deviation payoff} function is 
$u(\vec{\sigma}\mid t)$.
We use this unsubscripted lowercase $u$ throughout to denote a vector of deviation payoffs.
We define \term{regret} using deviation payoffs: 
\begin{equation}\epsilon(\vec{\sigma}) = \max_j u_j(\vec{\sigma}) - \sigma \cdot u(\vec{\sigma}).\label{eq:regret}\end{equation} 
A \term{symmetric Bayes-Nash Equilibrium} (BNE) is a symmetric profile $\vec{\sigma}$ such that no player has an incentive to deviate; equivalently, $\epsilon(\vec{\sigma}) = 0.$ 
As empirical games are themselves approximations, we focus on finding $\varepsilon$-BNE, which are symmetric profiles $\vec{\sigma}$ such that $\epsilon(\vec{\sigma}) \leq \varepsilon$, where $\varepsilon$ is small.


\paragraph{Parameterized Game Families} 
Let $\Gamma(v)$ denote the symmetric Bayesian game instance where the environment parameter $V$ takes value $v$. 
A \term{parameterized game family}, $\mathcal{G}(V)$, is the set of game instances $\{\Gamma(v)\mid v\in V\}$. 
In the game family, all payoff functions and functions that depend on payoffs (e.g., Eqs.~\ref{eq:ex_ante_exp_util}--\ref{eq:regret}) are parameterized by $V$.
Because regret depends on $V$, a profile $\vec{\sigma}$ that is an $\varepsilon$-BNE in one game instance may have regret larger than $\varepsilon$ in another game instance.
Simulator payoff samples take the form $\tilde{U}\big(s_j(t), \vec{s}(\vecsp{t}), t, \vec{t}, v)$.


\section{Learning Bayesian Game Families}
\label{sec:learn_game_families}

Simulator queries are costly, and the results are noisy due to randomness in strategies or in the game environment. 
We assume a fixed budget of simulator queries for learning and validation, so we must allocate queries across the parameter space, strategy space, and type space. 
This simulation data is used to train a model representing the deviation payoff function for a symmetric Bayesian game family.
We experimentally compare ex ante and interim game-family learning methods. 
The ex ante method becomes equivalent to the normal-form approach developed by \citet{gatchel2023learning} once types are abstracted away, and thus serves as our baseline in the Bayesian setting.

\subsection{Ex Ante Deviation Payoffs}
\label{sec:ex_ante}

We first train a neural network representing the ex ante deviation payoff function $\hat{u}: \Delta(S) \times V \to \mathbb{R}^{\abs{S}}$, 
where $\hat{u}_j(\vec{\sigma}, v)$ is the predicted payoff a symmetric player would receive by deviating to strategy $s_j$ when all other players play according to $\vec{\sigma}$ in game instance $\Gamma(v)$. 
Let $\nmix$ denote the number of $(\vec{\sigma}, v)$ pairs in our training set, let $\nobs$ denote the number of observations per pair, and let $\nsimq = \nmix \cdot \nobs$.
An \term{observation} for a given $(\vec{\sigma}, v)$ and deviation strategy $s_j$ involves sampling $\vec{s} \sim \vec{\sigma}$ and $(t, \vecsp{t})\sim \mu^p$, and querying the simulator for
$\tilde{U}\big(s_j(t), \vec{s}(\vecsp{t}), t, \vec{t}, v).$
A training example has the form $(\vec{\sigma}, v) \mapsto [\tilde{u}_j(\vec{\sigma}, v)]$, where $\tilde{u}_j(\vec{\sigma}, v)$ is the sample average of deviation payoffs across all observations for a given strategy $s_j$.
The \term{ground truth deviation payoff} is given by $\tilde{u}_j(\vec{\sigma}, v) = \frac{1}{\nobs}\sum_{i=1}^\nobs \tilde{U}(s_j(\idx{t}{i}), \idx{\vec{s}}{i}(\idx{\vec{t}}{i}), \idx{t}{i}, \idx{\vec{t}}{i}, v),$ where $(\idx{t}{i}, \idx{\vec{t}}{i}) \sim \mu^p$ and $\idx{\vec{s}}{i} \sim \vec{\sigma}$. 
We seek to minimize the training loss: 
\begin{equation}
\frac{1}{\nmix\cdot \abs{S}} \sum_{k=1}^m \sum_{s_j \in S} \Big(\tilde{u}_j(\vidx{\sigma}{k}, \idx{v}{k}) - \hat{u}_j(\vidx{\sigma}{k}, \idx{v}{k})  \Big)^2.
\label{eq:ex_ante_loss}
\end{equation}

\subsection{Interim Deviation Payoffs}

In Bayesian games, sampling over types represents a distinct source of noise in payoff estimates.
By conditioning estimates on the deviator's type, we can leverage type-specific information in each sample.  
We therefore propose learning the interim deviation payoff function, which is explicitly conditional on the symmetric deviator's type: $\hat{u}: \Delta(S) \times V \times T \to \mathbb{R}^{\abs{S}}$, where $\hat{u}(\vec{\sigma}, v \mid t)$ gives the predicted vector of conditional deviation payoffs, [$\hat{u}_j(\vec{\sigma}, v \mid t)$] for each $s_j \in S$ in game instance~$\Gamma(v)$.
An interim training example is a mapping $(\vec{\sigma}, v, \idx{t}{i}) \mapsto [\tilde{u}_j(\vec{\sigma}, v \mid \idx{t}{i})]$, where $\tilde{u}_j(\vec{\sigma}, v \mid \idx{t}{i})$ corresponds to the deviation payoff from a single observation.
We aim to minimize the training loss:
\[\frac{1}{\nsimq\abs{S}} \sum_{k=1}^\nmix \sum_{i=1}^\nobs \sum_{s_j \in S} \Big(\tilde{u}_j(\vidx{\sigma}{k}, \idx{v}{k} \mid \idx{t}{i}) - \hat{u}_j(\vidx{\sigma}{k}, \idx{v}{k} \mid \idx{t}{i})  \Big)^2.\]
We can compute ex ante deviation payoffs from the interim by marginalization: $\int_t \hat{u}_j(\vec{\sigma}, v \mid t) \mu(t)dt.$
In practice, this can be done via Monte Carlo integration by averaging predicted interim deviation payoffs over many sampled deviator types.
The marginalized interim deviation payoff vector for $(\vec{\sigma}, v)$ is approximated using $n$ Monte Carlo (MC) samples $\idx{t}{i}\sim \mu$ and the learned interim model:
\begin{equation*}
    \hat{u}(\vec{\sigma}, v) = \frac{1}{n} \sum_{i=1}^{n} \hat{u}(\vec{\sigma}, v \mid \idx{t}{i}).
\end{equation*}


\section{Using Learned Bayesian Game Families}
\label{sec:use_learned_game_fams}

\subsection{Deriving Equilibria}
Adapting existing techniques \citep{sokota2019learning, gatchel2023learning} to the Bayesian setting, we run an approximate Nash-finding algorithm on each game instance $\Gamma(v)$ using deviation payoffs predicted by the model: $[\hat{u}_j(\vec{\sigma}, v)]$ (ex ante) or $[\int_t \hat{u}_j(\vec{\sigma}, v \mid t) \mu(t)dt]$ (marginalized interim).
The resulting mixed-strategy profile is a \term{candidate $\varepsilon$-equilibrium}, $\vec{\sigma}_*$, if the predicted regret is at most $\varepsilon$; otherwise, the algorithm did not converge. 
We validate the candidate with a modest number of additional simulator queries to compute a high-fidelity  regret estimate of $\vec{\sigma}_*$ in $\Gamma(v)$; if this too is at most~$\varepsilon$, the approximate equilibrium is \term{confirmed}.
Running the algorithm from multiple starting points yields a set of equilibria suitable for any selection method.

\subsection{Empirical Mechanism Design (EMD)}
EMD is a motivating application for learned Bayesian game families.
Once trained, the game-family model can evaluate any game instance within---and often beyond---the trained range, supporting finer parameter searches than previous EMD approaches. 
When used in an optimization algorithm, it eliminates the need to train separate models at each iteration, reducing the algorithm's dependence on the sampling budget. 
See App.~\ref{app:use_game_fam_method_details} for pseudocode.\AAMASARXIVNOTE 


\subsection{Piecewise-Conditional Strategies}
\label{sec:piecewise}
In a BNE, no player can gain by deviating to another strategy in $S$. 
If $S$ includes all mappings from type to action, then an ex ante equilibrium is an interim equilibrium, as a player would also not wish to deviate conditional on its own type.
Given a restricted set $S$---the norm for empirical games---a player \emph{can} often benefit by deviating to a different strategy in~$S$ once its type is revealed. 
We consider a higher-order strategy that exploits such opportunities by selecting a base-level \term{atomic strategy} $s\in S$ conditional on revealed type~$t$.
We can approximate the optimal deviation by deriving a piecewise best response to $(\vec{\sigma}, v)$.
Let $\partition$ be a set of contiguous type intervals partitioning~$T$.
A \term{piecewise strategy} is a mapping $\pbr: \partition\to S$ that selects an atomic strategy for the interval containing~$t$. 

The learned interim model can approximate the ex ante expected payoff for playing a piecewise strategy in a given strategic context (e.g., where opponents play the BNE).
First, we collect a large set of samples $\typesamp$ from the type distribution~$\mu(T)$.
Next, we compute the deviation payoff vector conditioned on the deviator having a type in interval $\interval\in\partition$:
\begin{equation}
    \hat{u}(\vec{\sigma}, v \mid \interval) =  \frac{1}{\abs{\typesamp_C}}\sum_{\idx{t}{i}\in\typesamp_C} \hat{u}(\vec{\sigma}, v \mid \idx{t}{i}),
    \label{eq:cond_intvl_pred_pbr_payoff}
\end{equation}
where $\typesamp_\interval = \typesamp\cap\interval$.
The \term{piecewise best-response strategy}, $\pbr$, assigns each interval $\interval$ the best-response strategy conditioned on having type $t \in \interval$:
\begin{equation}
    \pbr(\interval) = \argmax_{s_j \in S} \hat{u}_j(\vec{\sigma}, v \mid \interval)
    \label{eq:piecewise_br}
\end{equation}
We can equivalently express $\pbr$ as a mapping from types to atomic strategies: for $t \in C$, $\pbr(t) \equiv \pbr(C)$.
Given $\hat{u}(\vec{\sigma}, v \mid \interval)$ and $\typesamp_\interval$ for each $\interval \in \partition$, the \term{predicted deviation payoff for playing the piecewise strategy} against $\vec{\sigma}$ is:
\begin{equation}
    \hat{u}_{\pbr}(\vec{\sigma}, v) = \frac{1}{\abs{\typesamp}}\sum_{\interval\in\partition} \abs{\typesamp_\interval}\cdot \hat{u}_{\pbr(C)}(\vec{\sigma}, v \mid \interval).
    \label{eq:pred_pbr_payoff}
\end{equation}
With $N$ simulator queries per $\vec{s}$, a high-fidelity estimate of the \term{true deviation payoff for playing the piecewise strategy} against $\vec{\sigma}$ is:
\begin{equation}
    \tilde{u}_{\pbr}(\vec{\sigma}, v) = \frac{1}{N}\sum_{\vec{s}\in\vec{S}}\Pr(\vec{s}\mid\vec{\sigma})\sum_{i=1}^N \tilde{U}(\pbr(\idx{t}{i}), \vec{s}(\idx{\vec{t}}{i}), \idx{t}{i}, \idx{\vec{t}}{i}, v). 
    \label{eq:true_pbr_payoff}
\end{equation}


\section{Dynamic Sponsored Search Auctions}
\label{sec:dynamic_auction}

Our case study evaluates game-family learning approaches for EMD in a dynamic sponsored search auction. 
A \term{sponsored search auction} allocates ad slots on a search page to advertisers based on their bids.
The publisher ranks bids based on the offered price---typically adjusted by factors such as click-through rates and advertiser quality---and allocates slots in descending order of effective bids.
Many studies model sponsored search as a one-shot simultaneous-move game \citep{lahaie2006analysis, edelman2007internet, varian2007position, thompson2009computational}, typically assuming perfect information on the basis that advertisers learn about competitors through repeated interactions.
When a bidder submits a bid, they may glean information based on the resulting state which may be used to adjust their bid. 
This dynamic adaptation, however, is abstracted away by one-shot models, which thereby do not capture imperfect adaptation or transient effects.

Our formulation aims to capture a simple form of the missing dynamics, by modeling a two-step bidding process. 
In the first stage, players simultaneously submit their initial bid based only on their valuation (type).
Each player is then informed about their tentative slot and its price, as well as the prices for other slots assuming competitor bids remained fixed. 
In the second stage, they may update their bid based on this observation.
Updates are submitted simultaneously, and each is received ``on time'' by the auction with a specified probability. 
The final bids determine the slot allocations and prices.
For example, a strategy may set the initial bid to valuation minus two and the final bid as its optimal response to the provisional state.
As all players have the option to change their bids, a player may sometimes be worse off by changing rather than keeping their initial bid. 
Because the updates are only probabilistically successful, players have an incentive to submit meaningful bids in the first stage.
This setup captures agents' imperfect information about opponents and their ability to gain information and adjust heuristically through interaction in a single-shot game.

We study a symmetric dynamic auction with 5 players, 4 ad slots, and 10 strategies.
The game family is parameterized by a \term{reserve requirement}~$\reserve$---the minimum effective bid to win a slot---typically set by the publisher. 
The mechanism, applied to final bids, is a weighted generalized second-price auction with quality-weighted reserves and Varian preference model~\citep{varian2007position}. 
Each advertiser has a quality score, $q$, and valuation, $\val$.
To participate, a bidder's \term{effective bid}---its quality-weighted bid---must meet the reserve. 
See App.~\ref{app:dynamic_game_details} for more details, including strategy specification. 


\section{Ex Ante vs Interim Learning}
\label{sec:ea_vs_i_learning}

We evaluate deviation-payoff performance for two learned models of the same Bayesian game family. 
The \term{ex ante} model inputs a symmetric mixed strategy $\vec{\sigma}$ and reserve~$\reserve$. 
The \textbf{interim} model also conditions on deviator quality score $q$ and valuation $\val$, and its predictions are marginalized over $(q, \val)$ for evaluation.  
With a training budget of $\nsimq$ simulator queries, we build six datasets satisfying $\nsimq = \nmix \cdot \nobs$, each containing $\nmix$ $(\vec{\sigma}, \reserve)$ pairs and $\nobs$ observations per pair; reserves range from 0.01 to 8. 
For each dataset, we train one ex ante neural network (NN) on $\nmix$ examples (payoffs averaged over the $\nobs$ observations) and one interim NN on all $\nmix\cdot\nobs$ observations. 
See App.~\ref{app:training_info} for training details.

\begin{figure*}[ht!]
\centering
\begin{tabular}{ccc}
    \includegraphics[width=\columnwidth]{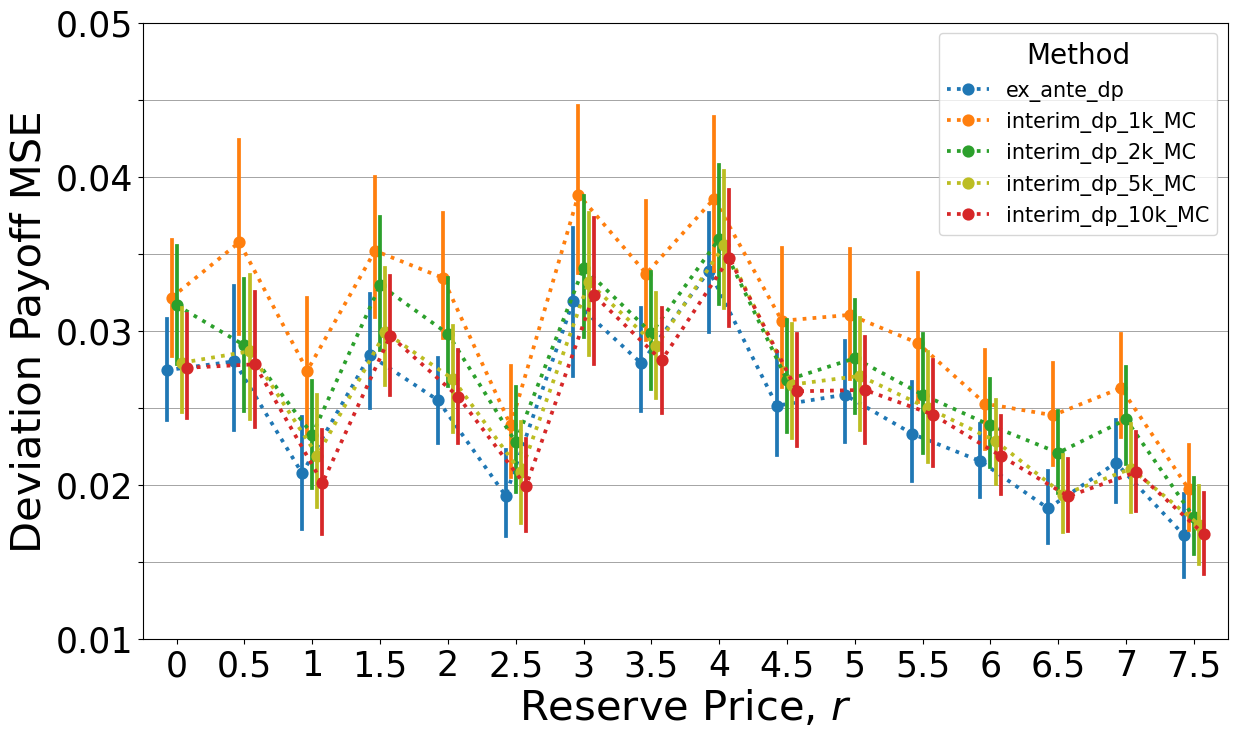} & 
    \includegraphics[width=\columnwidth]{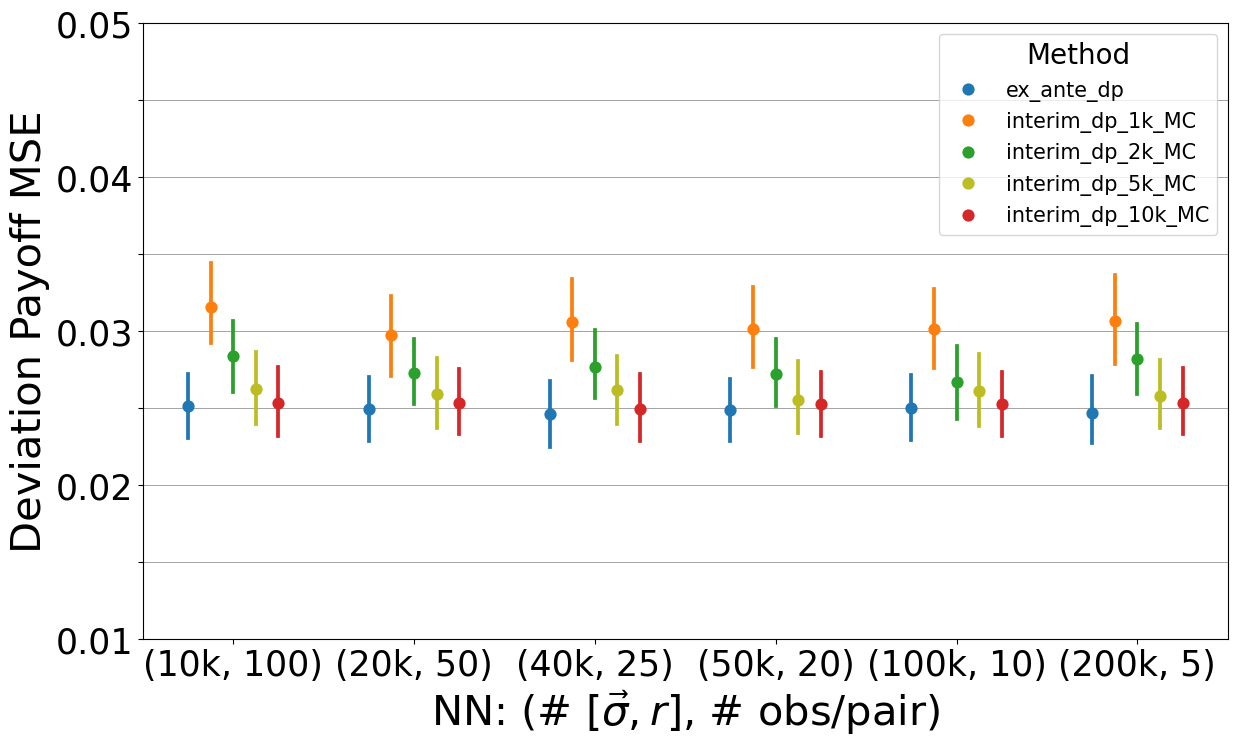} \\ 
    (a) & (b)
\end{tabular}
\caption{With sufficient marginalization samples, interim accuracy matches ex ante. 
Deviation payoff trends are consistent (a)~across the reserve range and (b)~across datasets.}
\label{fig:ml_test_results}
\end{figure*}

We use two test sets: a noisy, realistically sized dataset for a large empirical game family, and a larger, less noisy dataset for method validation.
Our results show that with sufficient marginalization samples, the interim model matches ex ante error, and exhibits better extrapolation. 
Performance trends on both test sets are consistent, suggesting the noisy test set provides reliable  insights.  

\subsection{Model Learning (ML) Test Results}

Fig.~\ref{fig:ml_test_results} compares deviation payoff errors among learned ex ante and interim game-family models. 
We compute the mean squared error (MSE) between ex ante or marginalized interim predictions and the ground truth using Eq.~\ref{eq:ex_ante_loss}; error bars show 95\% confidence intervals (CIs).
We compute marginalized interim deviation payoffs by averaging predicted interim payoffs over $\{\text{1k, 2k, 5k, 10k}\}$ Monte Carlo samples of deviator quality and valuation.
Plot~\ref{fig:ml_test_results}(a) shows errors across the reserve price range, with test-set reserve prices binned in intervals of width 0.5; performance is aggregated across the six trained models for each method.
The interim models with 5k and 10k Monte Carlo marginalization samples perform similarly to ex ante, with only a slight performance decrease for interim models with 2k and 1k marginalization samples. 
Plot~\ref{fig:ml_test_results}(b) shows model performance across training datasets, aggregated across the reserve-price range. 
Relative performance trends across the five methods are consistent across the six training datasets, suggesting the trends are robust to different $\nmix$ and $\nobs$ combinations.
For further discussion, see App.~\ref{app:ml_test}.

\subsection{Fine-Grained Grid Mixture Results}
\label{sec:extra_test_set}

\begin{figure}[htb]
    \centering
    \includegraphics[width=.96\columnwidth]{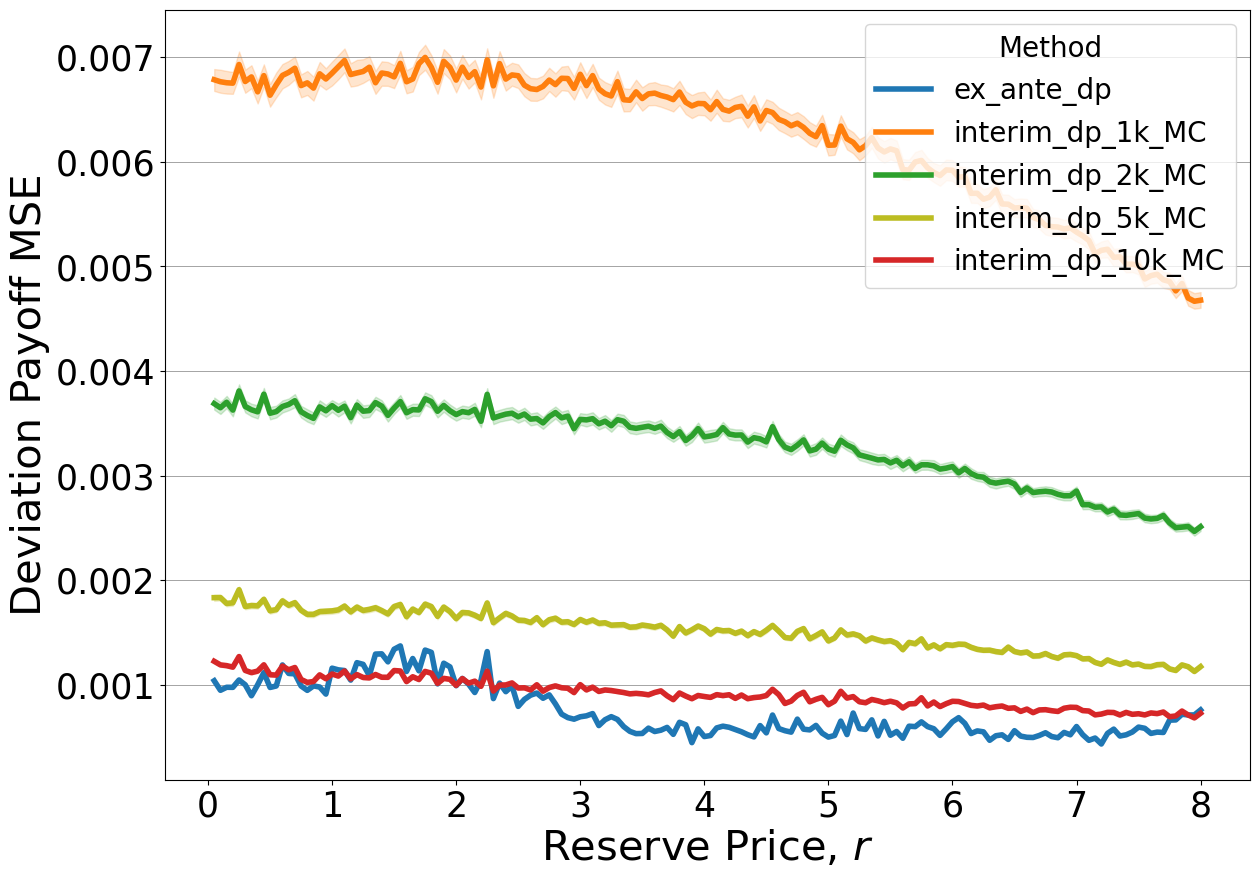}
    \caption{All models perform better on the fine-grained dataset than shown by the noisy ML test set (Fig.~\ref{fig:ml_test_results}), but with consistent trends.}
    \label{fig:extra_test_results_rleq8}
\end{figure}

\begin{figure}[ht!]
    \centering
    \includegraphics[width=.96\columnwidth]{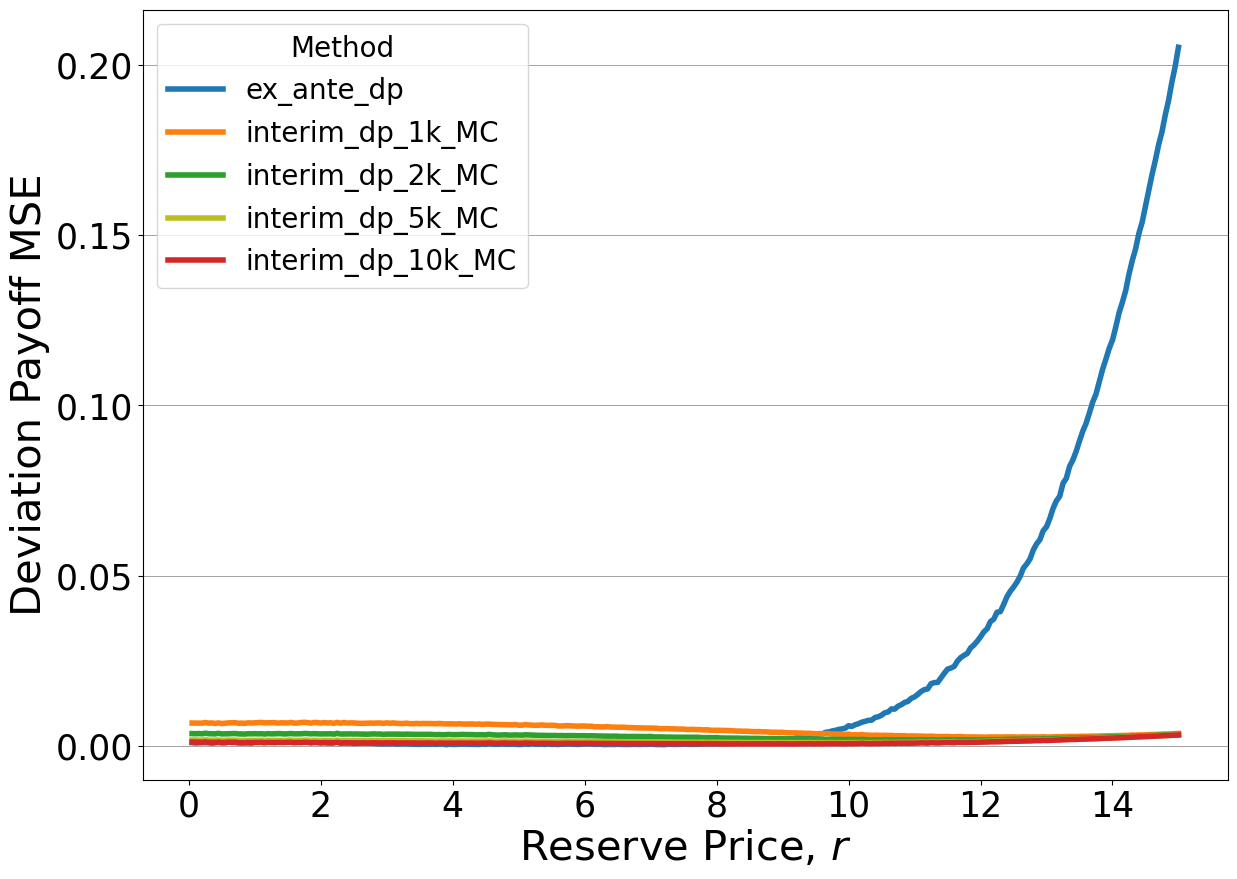}
    \caption{Interim but not ex ante models extrapolate well beyond the trained range, $\reserve \in (0,8].$}
    \label{fig:extra_test_results_rleq15_orig}
\end{figure}

For further evaluation, we construct a lower-noise dataset with 300 reserve prices ($r \in [0.05,15]$ in 0.05 steps), more mixed strategies per reserve, and more observations per mixture (see App.~\ref{app:extra_test}).
Figs.~\ref{fig:extra_test_results_rleq8}~and~\ref{fig:extra_test_results_rleq15_orig} show deviation-payoff MSEs (95\% CIs shaded) for ex ante and interim models trained on $\reserve \leq 8$, with errors computed as before. 
Performance is aggregated across the six NN models for each method. 
Fig.~\ref{fig:extra_test_results_rleq8} shows that errors are now smaller and more uniform across the reserve range, indicating that the learned models are more accurate than the ML test set captured. 
Relative performance is unchanged: interim models with 5k-10k Monte Carlo samples match ex ante accuracy; all errors are small relative to payoff scale (App.~\ref{app:training_info} Table~\ref{table:test_payoff_stats}).

As the parameter range covered by the training data must be set prior to any game-theoretic analysis, it may be too narrow, so extrapolation is beneficial for applications like EMD. 
Fig.~\ref{fig:extra_test_results_rleq15_orig} shows error across the reserve range, with $\reserve>8$ representing extrapolation results. 
All marginalized interim models extrapolate well, while ex ante error rises sharply in extrapolation.
These results suggest that interim models learn the relationship between $q$, $\val$, $\reserve$, and deviation payoff: at high reserves, only bidders with $q \cdot \val \geq \reserve$ can substantively participate, and those few can secure large deviation payoffs by shading their bids. 
This extrapolation capability is a key advantage of the interim learning approach.

To examine extrapolation performance over different ranges, we train additional neural network models on subsets of the original training data.
We train both ex ante and interim models using only samples with $\reserve \leq \maxr$, for $\maxr \in \{1,\dots,7\}$ and each of the six initial training datasets (2 functions $\times$ 6 datasets $\times$ 7 cutoffs = 84 total NNs). 
All models use the same architecture and hyperparameter settings that were identified as optimal for their respective payoff representation under the full training dataset. 
We evaluate models on the fine-grained test set from Figs.~\ref{fig:extra_test_results_rleq8} and~\ref{fig:extra_test_results_rleq15_orig}.

\begin{figure}[ht]
    \centering 
    \includegraphics[width=.96\columnwidth]{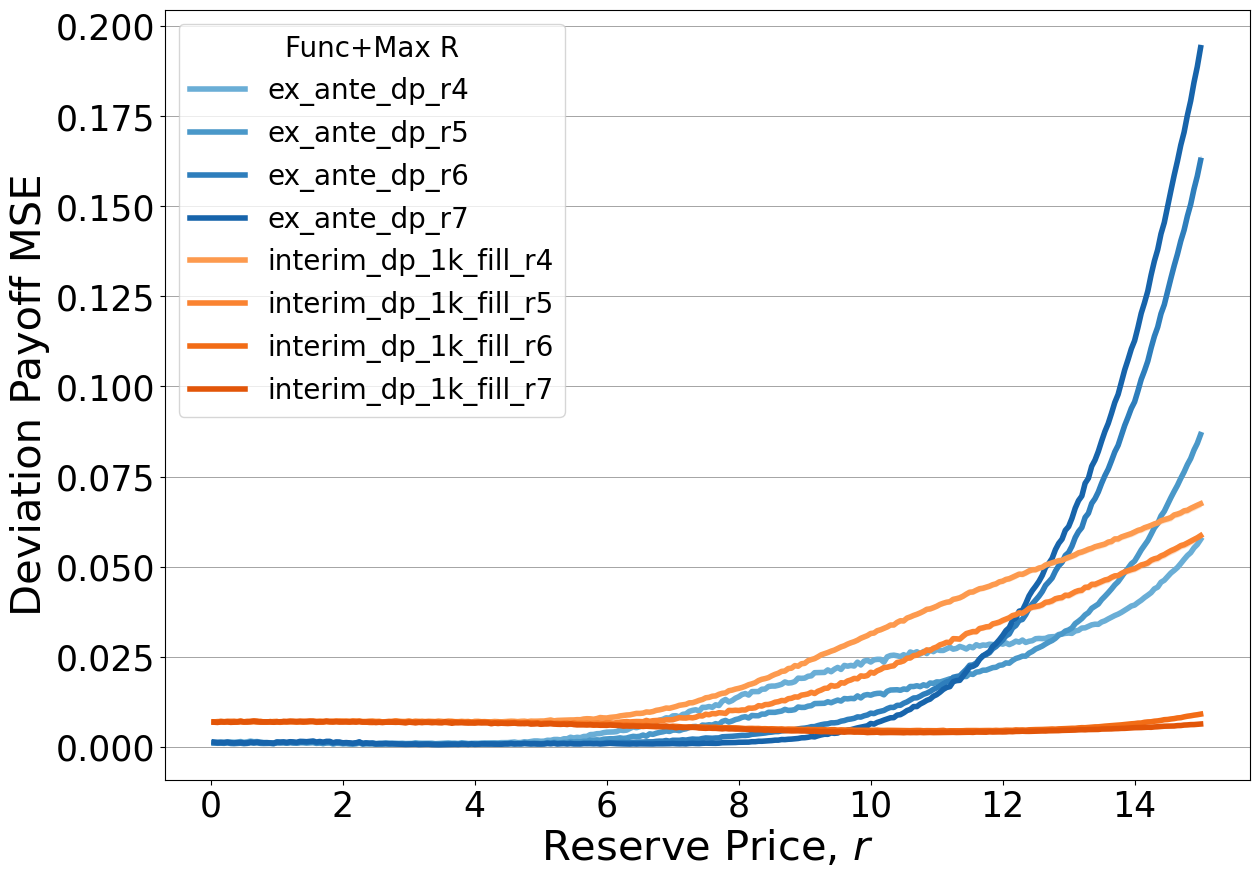}
    \caption{Extrapolation from training over various ranges.
    For training conditions most similar to those the models were tuned on, only interim models maintain strong extrapolation performance. With smaller datasets and more restricted ranges, ex ante and interim models perform comparably.}
    \label{fig:extra_extrapolation_results_maxr_geq4}
\end{figure}

Fig.~\ref{fig:extra_extrapolation_results_maxr_geq4} shows performance for models trained on datasets with $\maxr \in \{4,\dots,7\}$. 
Each curve aggregates results across six neural networks, with confidence intervals shaded.
Under training conditions most similar to the original experiment ($\maxr\in\{6,7\}$), interim models continue to extrapolate effectively, whereas ex ante models do not. 
For $\maxr\in\{4, 5\}$, extrapolation performance is comparable between the two methods.
This difference is likely affected by the smaller filtered datasets and fixed hyperparameters---each unit decrease in $\maxr$ reduces the training dataset size by roughly 12.5\% relative to the original training set for which the architecture and hyperparameters were tuned. 
It is also plausible that the interim method benefits from learning conditional payoffs over a broader range, where the relationship between environment parameter, type, and payoff is more apparent.
At lower maximum-reserve cutoffs, both methods exhibit expected degradation in extrapolation performance due to the limited training data. 
Together with Fig.~\ref{fig:extra_test_results_rleq15_orig}, these results suggest that when trained on ample data across a sufficiently wide parameter range, interim---but not ex ante---models demonstrate robust and consistent extrapolation.
App.~\ref{app:extra_test} provides statistics for each filtered dataset, and shows ex ante and interim results for all $\maxr$ values.


\begin{figure}[h!]
\centering
\begin{tabular}{c}
    \includegraphics[scale=.2]{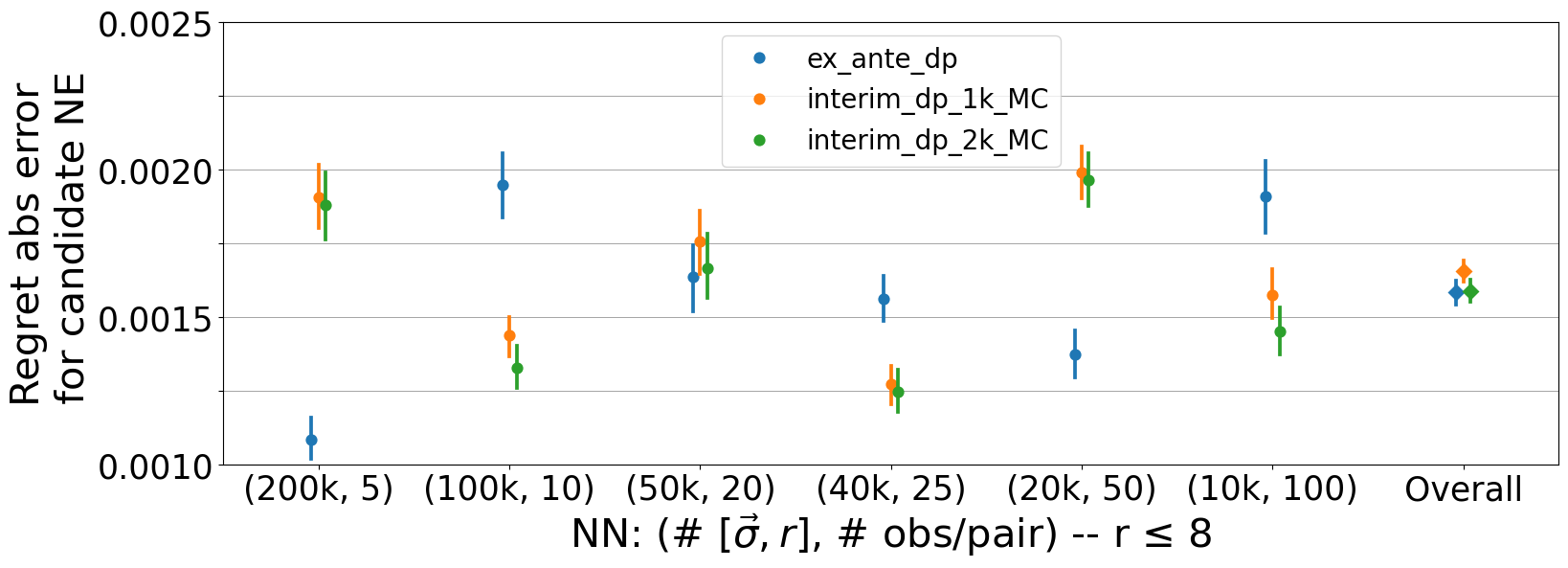} \\  
    \includegraphics[scale=.2]{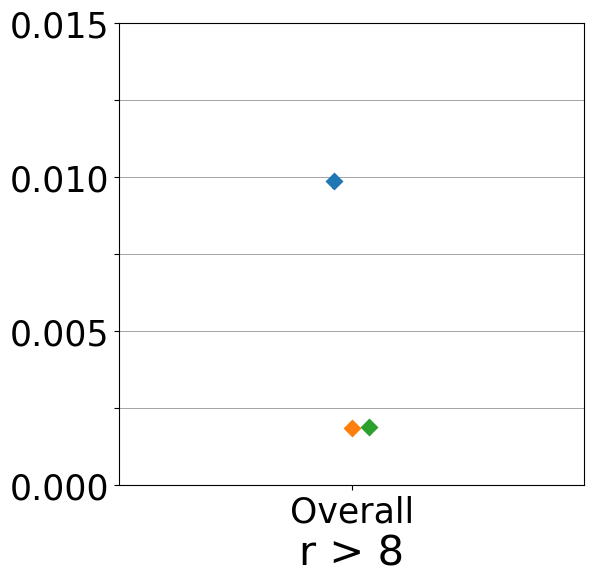} 
\end{tabular}
\caption{On the trained range ($\reserve \leq 8$), the interim method has regret error comparable to ex ante. In extrapolation, the interim method has lower error than ex ante.}
\label{fig:cand_ne_regr_abs_err}
\end{figure}

\section{Using the Learned Bayesian Game-Family Model}
\label{sec:use_learned_model_experiments}

\subsection{Deriving Equilibria}
\label{sec:find_appx_ne} 

To derive equilibria for each of the 300 game instances used in \S\ref{sec:extra_test_set}, we separately run replicator dynamics (RD) \citep{taylor1978evolutionary} with each learned model $\hat{u}$ from fixed initial points.
Fig.~\ref{fig:cand_ne_regr_abs_err} shows the mean absolute error between predicted and true\footnote{Throughout, ``true'' payoff or regret refers to a high-fidelity simulated estimate.} regret across all candidate $0.01$-BNE returned by RD using each NN (circles) and overall (diamonds) for the trained range (top) and in extrapolation (bottom). 
Both methods achieve low regret error for candidate equilibria on the trained range, but the ex ante method is unreliable in extrapolation, with regret error around $\varepsilon=0.01$.

Recall that $\hat{\epsilon}(\vec{\sigma})$ refers to predicted regret and $\epsilon(\vec{\sigma})$ denotes (a high-fidelity estimate of) the true regret.
We classify mixed-strategy profiles returned by RD into one of four categories:
\begin{center}
\begin{tabular}{c|c}
    Category & Description \\ 
    \hline 
    Did Not Converge & $\hat{\epsilon}(\vec{\sigma}) > \varepsilon$, mixture discarded \\
    Dead Mixture & $\hat{u}$ is $\vec{0}$, killing Nash-finding \\
    Rejected Candidate Mixture & $\hat{\epsilon}(\vec{\sigma}) \leq \varepsilon$; $\epsilon(\vec{\sigma}) > \varepsilon$ \\
    Confirmed Candidate BNE & $\hat{\epsilon}(\vec{\sigma}) \leq \varepsilon$; $\epsilon(\vec{\sigma}) \leq \varepsilon$
\end{tabular}
\end{center}
Table~\ref{tab:nash_stats} reports statistics on the percentage of rejected candidate mixtures, percentage of dead mixtures, and number of holes, separated by game instances within trained range ($r \leq 8)$ and in extrapolation ($r > 8$).
A hole occurs when there are no confirmed BNE approximated by the model for a given game instance. 
Consistent with Fig.~\ref{fig:cand_ne_regr_abs_err}, both methods perform well on the trained range, with a candidate mixture rejection rate below 3\% across all neural network models. 
In extrapolation, however, between 40\% and 50\% of mixtures derived by an ex ante model are rejected.
Furthermore, for approximately 9\% of RD runs with ex ante models, the predicted payoff vector consists entirely of  zeros, effectively ``killing'' the RD update process. 
Collectively, this results in substantial holes in the expected revenue curve: ex ante revenue curves have \emph{at least} 51 holes, and for 51 game instances (17\%) \emph{none} of the six ex ante models identify an equilibrium that is confirmed. 
Additional details and full RD mixture classification results are provided in App.~\ref{app:nash_approximation}.

\begin{table}[ht]
\centering
\caption{Summary of RD mixture classification results.}
\label{tab:nash_stats}
\begin{tabular}{llcc}
\toprule
\textbf{Range} & \textbf{Metric} & \textbf{Ex Ante} & \textbf{Interim (1k MC)} \\
\midrule
\multirow{3}{*}{$r \le 8$} 
 & \textbf{Rejected NE} &  &  \\
 & \hspace{1em}Min & 0.51\% & 0.51\% \\
 & \hspace{1em}Avg & 1.58\% & 1.52\% \\
 & \hspace{1em}Max & 2.84\% & 2.50\% \\
\midrule
\multirow{12}{*}{$r > 8$}
 & \textbf{Rejected NE} &  &  \\
 & \hspace{1em}Min & 39.87\% & 0.00\% \\
 & \hspace{1em}Avg & 43.71\% & 1.85\% \\
 & \hspace{1em}Max & 50.00\% & 4.29\% \\
 & \textbf{Dead Mixture} &  &  \\
 & \hspace{1em}Min & 0.71\% & 0.00\% \\
 & \hspace{1em}Avg & 8.69\% & 0.00\% \\
 & \hspace{1em}Max & 15.71\% & 0.00\% \\
  & \textbf{Num Holes} &  &  \\
 & \hspace{1em}Min & 51 & 0 \\
 & \hspace{1em}Avg & 69.17 & 2.17 \\
 & \hspace{1em}Max & 77 & 6 \\
\bottomrule
\end{tabular}
\end{table}


\subsection{Application to Empirical Mechanism Design}

Our objective is to identify the reserve price(s) that maximize expected revenue in equilibrium.
Game-family learning facilitates the approximation of multiple, distinct BNE, improving the robustness of statistics computed based on the $\varepsilon$-BNE set (e.g., worst case, average, max entropy).
While the appropriate equilibrium selection method is domain-dependent, we found that relying on the first-returned equilibrium was unreliable, as neighboring game instances sometimes had different equilibria, affecting revenue smoothness.

\subsubsection{Grid Optimization}
\begin{figure}[ht!] 
\centering
\begin{tabular}{c}
\includegraphics[width=.96\columnwidth]{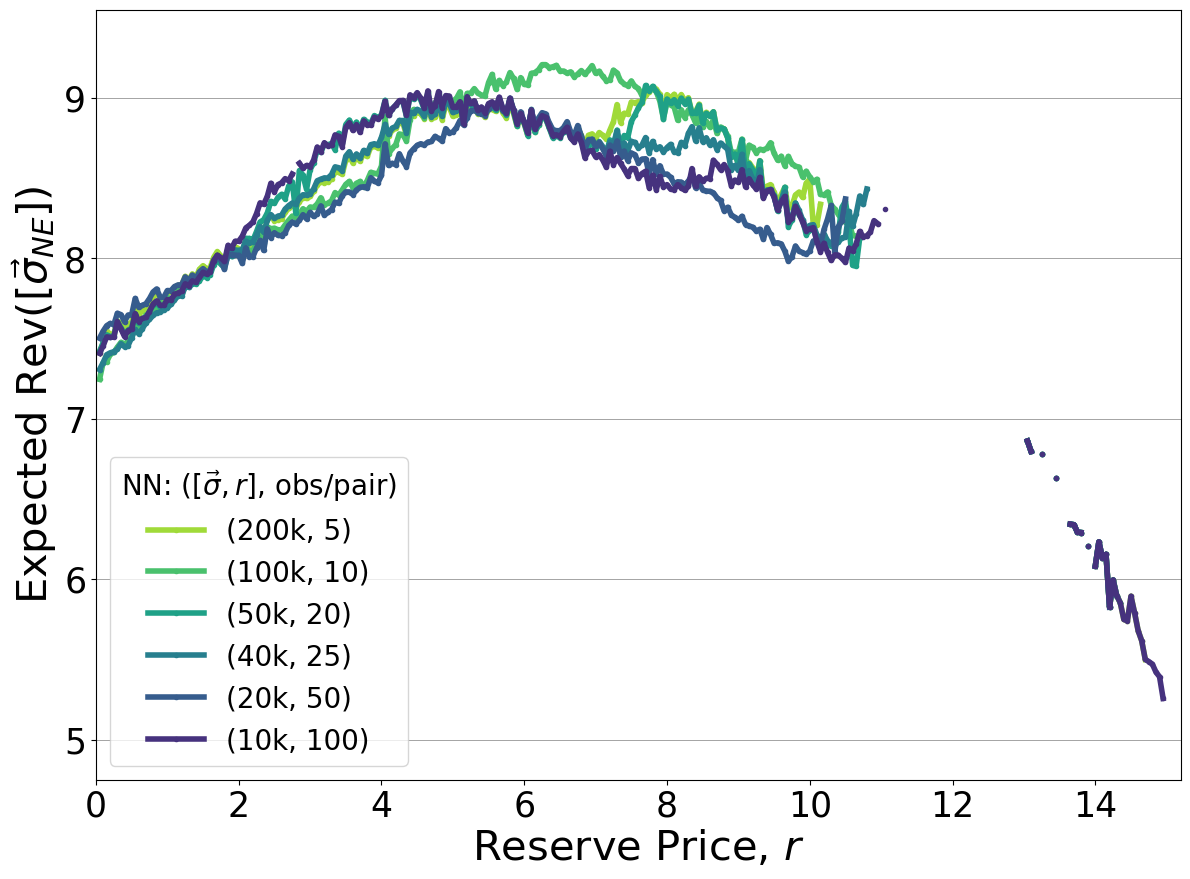} \\
(a) \\
\includegraphics[width=.96\columnwidth]{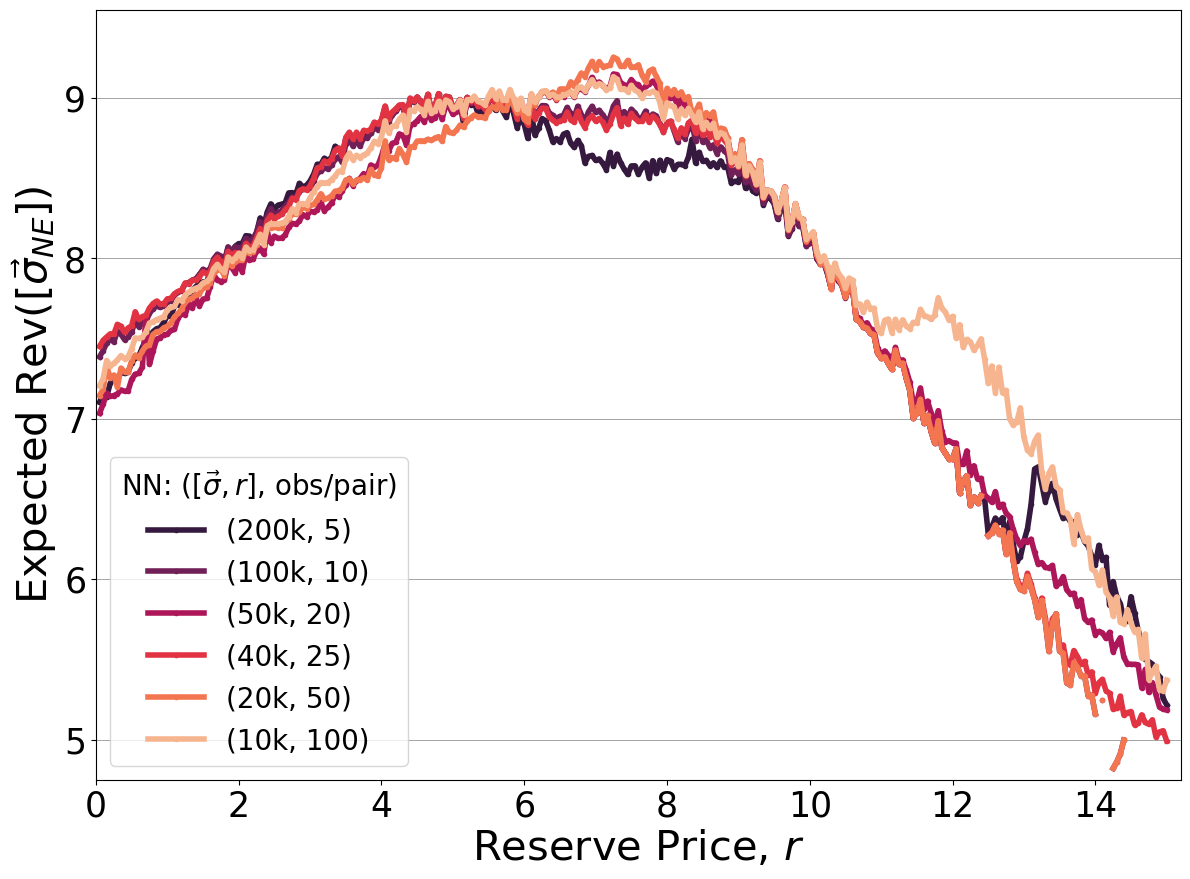} \\ 
(b)  \\ 
\end{tabular}
\caption{Expected revenue for equilibria derived by (a)~ex ante and (b)~interim models.}
\label{fig:emd_grid_optimization}
\end{figure}

We first conduct a grid search on the average expected revenue curve using the $\varepsilon$-BNE sets from \S\ref{sec:find_appx_ne}.
Fig.~\ref{fig:emd_grid_optimization} shows expected revenue in equilibrium across all confirmed equilibria found by each (a)~ex ante and (b)~interim model. 
The overall shape of revenue curves is consistent across models, demonstrating the effectiveness of conducting EMD with a learned game family.
For some game instances, the curves diverge: each learned model may derive distinct---yet still valid---equilibria, and small shifts in the set of equilibria derived can translate into noticeable differences in the objective. 
Training multiple models (on same or different data) and evaluating the objective on the full set of equilibria derived may address this challenge.
Moreover, the presence of multiple equilibria and the possible need for extrapolation suggest that learning the objective function directly would be insufficient.


In extrapolation, the ex ante curves exhibit significant gaps where every candidate equilibrium is rejected.
Relying solely on an ex ante revenue curve leaves the possibility of a second revenue peak around $\reserve=11.$
In contrast, with any interim model we have confidence that the optimal reserve is below 9, and interim extrapolation is crucial for this confirmation.
Although no single optimal reserve price emerges, identifying the optimal revenue plateau (e.g., $6\leq \reserve \leq 8$) remains valuable as optimal plateaus have been present in past EMD studies \citep{brinkman2017empirical}.
Worst-case expected revenue results are provided in App.~\ref{app:emd}.

\subsubsection{Local Search Optimization}
We compare using ex ante and interim models in stochastic hill climbing and simulated annealing local search algorithms (described in App.~\ref{app:use_game_fam_method_details}).
Stochastic hill climbing terminates when all uphill neighbors have been explored ($\reserve \pm 0.05, \reserve \pm 0.1, \reserve \pm 0.25)$ or after 50 iterations. 
Simulated annealing terminates after 50 iterations, but typically evaluates a comparable number of game instances due to initial high temperature causing repeated selections.
To evaluate a given game instance with learned model $\hat{u}$, we run RD with $\hat{u}$ (using 1000 Monte Carlo type samples for interim) starting from the same initial mixtures as in the previous section.  
We use model~$\hat{u}$ to compute predicted regret, and identify candidate equilibria. 
Expected revenue for a candidate equilibrium $\vec{\sigma}$ is approximated by sampling 100 pure profiles from $\vec{\sigma}$, and computing average revenue across 10,000 auction settings for each pure profile.
The expected revenue in equilibrium is the average revenue across all \textit{candidate} equilibria.  
After search termination, we identify the game instance with the highest revenue in equilibrium, and use true deviation payoff samples to determine the set of confirmed equilibria. 
The final reported expected revenue in equilibrium is the average revenue over all \textit{confirmed} equilibria. 

\begin{figure}
    \centering
    \includegraphics[width=\columnwidth]{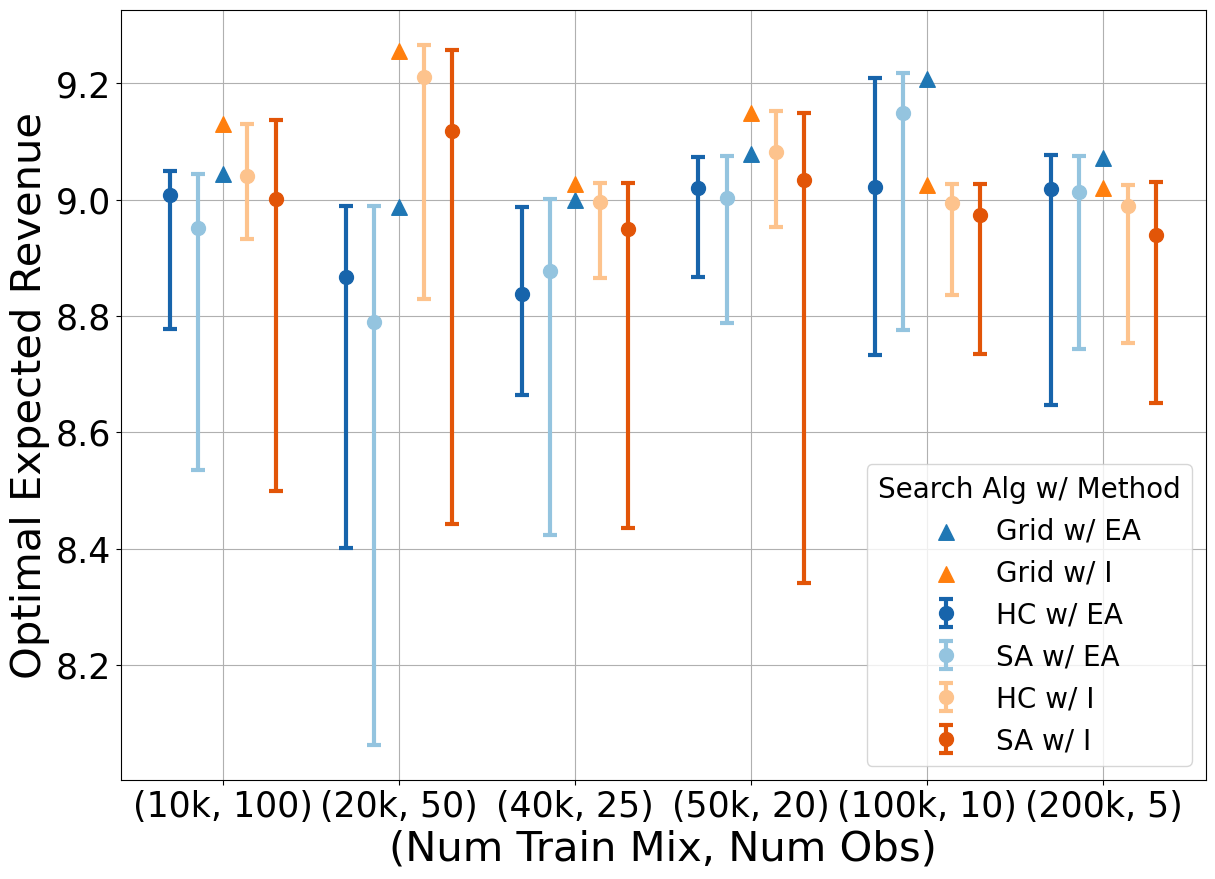} 
    \caption{Local search with learned game families and limited restarts identifies high-revenue game instances comparable on average to grid optimization. }
    \label{fig:local_search_results}
\end{figure}

Fig.~\ref{fig:local_search_results} shows 95\% bootstrap confidence intervals for optimal expected revenue in equilibrium from hill climbing (HC) and simulated annealing (SA) using ex ante (EA) or interim (I) models. 
For each combination, we perform 50 random $\reserve$ restarts for robust evaluation.
As 5 restarts is more practically realistic, we sample 5 (of~50) restart experiments with replacement, and determine the game instance with the max revenue for confirmed equilibria. 
We repeat this process 100,000 times, and plot 95\% confidence intervals around the mean optimal expected revenue for confirmed equilibria.  
For each model, we also plot the maximum expected revenue based on grid optimization. 
In nearly all cases, the upper confidence bound aligns with the optimal value from grid optimization. 
Note that Nash approximation with ex ante models is deterministic whereas Nash approximation with interim is not, which is why the interim confidence interval may extend beyond the max revenue from grid optimization. 
The decrease in max revenue from grid optimization to average max revenue in local search is less than 2.25\%. 
This is notable considering the granularity ($\delta = 0.05$, $0.05 \leq \reserve \leq 15$) and that local search evaluates only 30-50\% of the game instances on average compared to grid optimization (App.~\ref{app:emd} Fig.~\ref{fig:num_game_instances_evaluated}). 
This experiment highlights the effectiveness of local search with learned game families and a modest number of restarts.

\subsection{Piecewise-Conditional Strategies}
\label{sec:computing_pwbr_results}

Without additional simulation or training, we use the learned interim model to (1)~compute a piecewise best-response strategy to an $\varepsilon$-BNE, and (2)~quantify the payoff gain by playing this strategy while opponents play the $\varepsilon$-BNE. 
We reduce the type space to a single dimension by using $q\cdot \val$, which represents a deviator's maximum effective bid.
We partition this type space into $\abs{\mathcal{C}} = 5$ equiprobable intervals.
For a given $\reserve$ and confirmed $\vec{\sigma}_{\BNE}$, we use the learned interim model to compute $\pbr$ according to Eq.~\ref{eq:piecewise_br} with $\abs{\typesamp}= 100,000$ sampled values of $q$ and $\val$.
The predicted deviation gain from $\vec{\sigma}_{\BNE}$ to $\pbr$ is $\hat{u}_{\pbr}(\vec{\sigma}_{\BNE}, \reserve) - \vec{\sigma}_{\BNE} \cdot \hat{u}(\vec{\sigma}, \reserve)$, where $\hat{u}_{\pbr}$ is given by Eq.~\ref{eq:pred_pbr_payoff} and $\hat{u}$ is the marginalized interim deviation payoff vector.
The corresponding high-fidelity gain, $\tilde{u}_{\pbr}(\vec{\sigma}_{\BNE}, \reserve) - \vec{\sigma}_{\BNE}\cdot \tilde{u}(\vec{\sigma}_{\BNE}, \reserve)$, uses Eq.~\ref{eq:true_pbr_payoff} for $\tilde{u}_{\pbr}$.

\begin{figure}[ht]
\centering
    \includegraphics[width=\columnwidth]{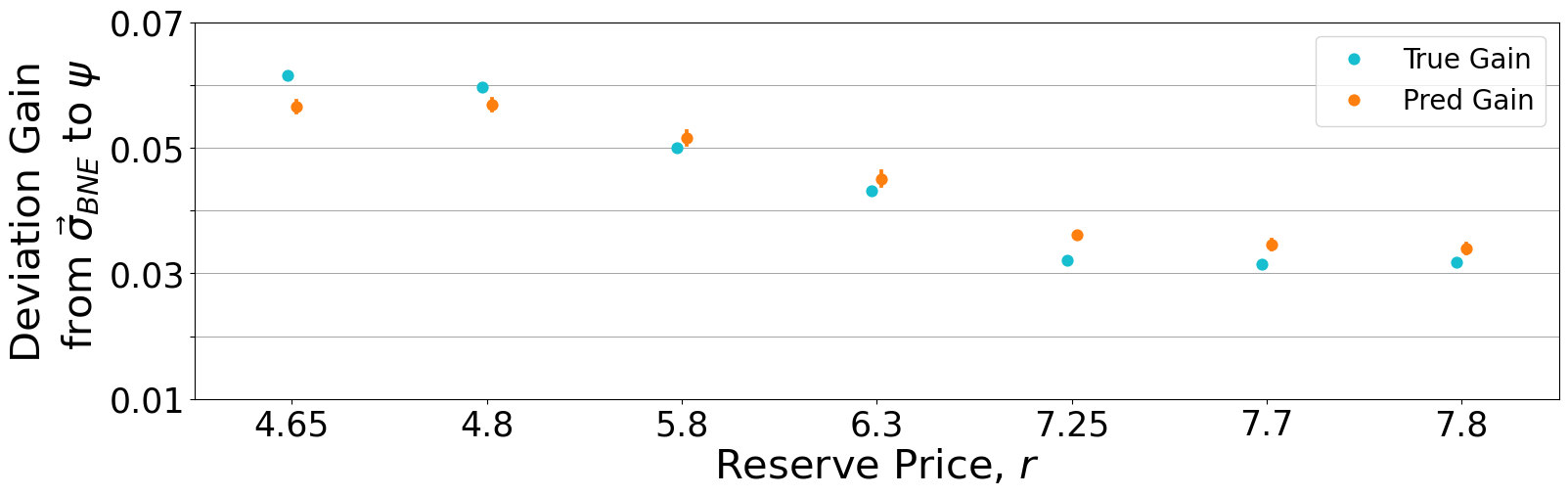}
    \caption{A player can gain payoff larger than $\varepsilon = 0.01$ by playing a piecewise best response ($\pbr$) to an $\varepsilon$-BNE compared to playing the $\varepsilon$-BNE.}
    \label{fig:piecewise_gain}
\end{figure}

Fig.~\ref{fig:piecewise_gain} compares predicted and true gains for playing $\pbr$ in response to equilibria in optimal game instances identified via grid optimization. 
For each $\reserve$, interim NN, and confirmed $\varepsilon$-BNE, we compute $\pbr$ and its corresponding predicted and true payoff gains; each point shows the mean gain across all 6 NNs and the equilibria they derived.
The gains from deviation entail a corresponding regret for the profile they are deviating from in a game with the  $\pbr$ added to $S$.
In other words, the derived BNE candidates have been refuted as 0.01-BNE.
Extending the learned Bayesian game family model to include these composite strategies would be needed to derive new  equilibria and optimal mechanism parameters in the augmented game family. 

\section{Conclusion} 
Through an in-depth EMD study of a dynamic sponsored search auction, we demonstrate the advantages of learning an interim model for Bayesian game families.
It achieves low payoff error across the trained parameter range and in extrapolation, and is overall more reliable at approximating Bayes-Nash equilibria than the ex ante model. 
Its extrapolation capability enables confident identification of the optimal reserve range via grid optimization. 
Finally, the model supports the computation of piecewise best-response strategies without additional sampling, which may effectively guide expansion of the strategy set.

\balance
\bibliographystyle{ACM-Reference-Format} 
\bibliography{references}



%% file: appendix.tex
\newpage
\appendix
\onecolumn

\section*{Appendix Organization}
\begin{itemize}
    \item App.~\ref{app:method_details} contains supplemental information for Section \ref{sec:learn_game_families} Learning Bayesian Game Families and Section~\ref{sec:use_learned_game_fams} Using Learned Bayesian Game Families. 
    \item App.~\ref{app:dynamic_game_details} provides additional information about the dynamic sponsored search auction described in Section~\ref{sec:dynamic_auction} Dynamic Sponsored Search Auctions.
    \item App.~\ref{app:game_learning_details} describes NN training information and provides additional results and discussion for Section~\ref{sec:ea_vs_i_learning} Ex Ante vs Interim Learning. 
    \item App.~\ref{app:use_learned_model} gives more information about experiment setup and additional results for experiments described in Section~\ref{sec:use_learned_model_experiments} Using the Learned Bayesian Game-Family Model.
\end{itemize}

\section{Methodological Details}
\label{app:method_details}

\subsection{Learning Bayesian Game Families}
\label{app:learn_method_details}

\paragraph{Interim Loss Function.} 
In principle, we could define $\tilde{u}_j(\vec{\sigma}, v \mid \idx{t}{i})$ as the sample average of deviation payoffs over many $\vec{t}$. 
Since our goal is to compare ex ante and interim learning approaches, for fair comparison we use the same training data (i.e., underlying $\tilde{U}$ payoff samples) with a single $t$ and $\vec{t}$ sampled for each observation, which is the natural approach for ex ante learning. 
It is possible that the alternative interim ground truth approach might benefit interim learning specifically, though it is not obvious how many $\vec{t}$ samples per $t$ to employ, given the tradeoffs resulting from a fixed budget of simulator queries.

\paragraph{Interim Deviation Payoff Error.} 
The deviation payoff MSE for a given interim model $\hat{u}$ with $n$ marginalization samples is equal to the squared error between the ground truth (ex ante) deviation payoff vector and the marginalized interim deviation payoff vector, averaged across all strategies and $(\vec{\sigma}, v)$ pairs in the dataset. 

\subsection{Using Learned Bayesian Game Families}
\label{app:use_game_fam_method_details}




\subsubsection{Empirical Mechanism Design}

Algorithm~\ref{alg:local_search_emd} shows how the learned Bayesian game-family model may be used in a general local search algorithm.  
For some initial parameter $v$, we use the learned model to find at least one $\varepsilon$-BNE, denoted by $\vec{\sigma}_*$, in game instance $\Gamma(v)$.
Given a set of $\varepsilon$-BNE in $\Gamma(v)$, we apply the pre-specified equilibrium selection or aggregation method and compute the equilibrium objective value, $z_*$.
Then we select some neighbor $v'$, derive equilibria in $\Gamma(v')$, and and compute objective function information, $(\vec{\sigma}'_*, z'_*)$, for game instance $\Gamma(v')$.
According to the local search algorithm and the values $z_*$ and $z'_*$, we determine which game instance to explore next.
In general, if $z'_*$ is better than $z_*$, then we accept neighbor $v'$ as our next parameter and relabel as $v$. 
In the case of simulated annealing, for example, the probability of accepting a game instance with a worse objective value in equilibrium decreases during subsequent iterations, as determined by the temperature cooling schedule. 
This algorithm can easily be extended to look at multiple neighbors at each iteration. 
In stochastic hill climbing, the algorithm may select a neighbor $v'$ with probability proportional to the relative improvement in $z'_*$ compared to the other neighboring game instances. 
This process is repeated for a pre-specified number of iterations. 
The mechanism designer selects the parameter setting where the induced game instance has the best equilibrium objective value out of the set of game instances evaluated.

\begin{algorithm}[ht]
\DontPrintSemicolon
\caption{EMD with a general local search algorithm using a learned Bayesian game family.}
\label{alg:local_search_emd}
\KwIn{$\hat{u}$, \textit{numIters}}
\Begin{
    $v \gets \operatorname{getInitParam()}$\;
    $(\vec{\sigma}_*, v) \gets \operatorname{findNash}(\hat{u}, \text{param}=v)$\;
    $z_* \gets \operatorname{ComputeNEObjective}(\vec{\sigma}_*, v)$\;
    $\text{history} \gets \{(v, z_*)\}$\;
    
    \Repeat{numIters}{
        $v' \gets \operatorname{getNeighborParam}(v)$\;
        $(\vec{\sigma}'_*, v') \gets \operatorname{findNash}(\hat{u}, \text{param}=v')$\;
        $z_*' \gets \operatorname{ComputeNEObjective}(\vec{\sigma}'_*, v')$\;
        $(v, z_*) \gets \operatorname{getNextState}((v, z_*), (v', z_*'))$\;
        $\text{history}[v] \gets z_*$\;
    }
    
    \Return $\displaystyle \arg\max_{v} \,\text{history}[v]$\;
}
\end{algorithm}

    
    

\section{Dynamic Game Details}
\label{app:dynamic_game_details}

Capturing the missing dynamics abstracted away in classic one-shot perfect-information sponsored search auctions, we introduce a two-stage bidding dynamic game that allows bidders to attempt to adjust to the provisional outcome they observe before the auction clears. 
To our knowledge, no prior work adopts this particular two-stage structure.
The most closely related studies examine other sequential settings: \citet{vorobeychik2008equilibrium} find pure equilibria in a symmetric sequential auction game with four greedy bidding strategies and 
\citet{nisan2011bestresponse} study best-response auctions where agents sequentially best respond until convergence; then the auction executes.
We cast the two-stage auction as a Bayesian game, preserving some tractability while still modeling the bid adaptation dynamics. 

In the auction game family, each player's type consists of two components: their valuation per click, $\val \sim U(0, 25)$, and their quality score, $q \sim U(0, 1)$.
We use the Varian preference model \citep{varian2007position}.
Specifically, the click-through rate (CTR) matrix is separable into a publicly known fixed vector that denotes the CTR for each advertising slot and the (private) quality score vector. 
So, the value at index $(j, i)$ of the CTR matrix gives the probability that advertiser $i$'s ad is clicked in slot $j$, and is equivalent to the product of the CTR for slot $j$ and player $i$'s quality score $q$. 
Bids are integers, and we assume that no bidder will bid above their valuation. 
To be eligible to win a slot, a bidder's effective bid, the product between their quality score and submitted bid, must meet the reserve requirement. 

\subsection{Strategy Specification}
For the initial bid, a strategy specifies bidding $\max(\left\lfloor \val \right\rfloor - \mathit{offset}, 0)$, for some integer offset. 
For the final bid, a strategy specifies whether to submit a \term{best response bid}, an updated bid representing the optimal value based on the tentative slot allocations and prices.
We conducted significance tests to confirm that there were statistically significant differences in expected payoffs for each pair of strategies across the reserve price range. 
In our experiments, we consider a game family with five offset increments, \{0, 2, 4, 6, 8\}, and consider strategies that involve submitting a best response bid update or not.  
The strategy set is characterized by: 

\begin{table}[ht]
\centering 
\begin{tabular}{c|c|c}
$S$ & Initial Bid & Final Bid \\ 
\hline 
$s_0$ & $\max(\left\lfloor \val \right\rfloor, 0)$ & No change \\
$s_1$ & $\max(\left\lfloor \val \right\rfloor -2, 0)$ & No change \\
$s_2$ & $\max(\left\lfloor \val \right\rfloor -4, 0)$ & No change \\
$s_3$ & $\max(\left\lfloor \val \right\rfloor -6, 0)$ & No change \\
$s_4$ & $\max(\left\lfloor \val \right\rfloor -8, 0)$ & No change \\
$s_5$ & $\max(\left\lfloor \val \right\rfloor, 0)$ & Best response bid \\
$s_6$ & $\max(\left\lfloor \val \right\rfloor - 2, 0)$ & Best response bid \\
$s_7$ & $\max(\left\lfloor \val \right\rfloor -4, 0)$ & Best response bid \\
$s_8$ & $\max(\left\lfloor \val \right\rfloor -6, 0)$ & Best response bid \\
$s_{9}$ & $\max(\left\lfloor \val \right\rfloor -8, 0)$ & Best response bid \\
\end{tabular}
\end{table}

\newpage


\section{Dyamic Auction Game-Family Learning Details \& Results}
\label{app:game_learning_details}

\subsection{Training \& Evaluation Information}
\label{app:training_info}

\subsubsection{Datasets}
We assume a budget of $\nsimq = 1000000$ total training queries, and construct six training sets with different values of $\nmix$ $(\vec{\sigma}, v)$ pairs and $\nobs$ observations per pair. 
Recall that for a given $(\vec{\sigma}, v)$ pair, the ex ante target is the deviation payoff vector averaged over the $\nobs$ observations.
For a given training set, the interim model receives $\nmix\cdot \nobs$ training examples and the ex ante model receives $\nmix$ training examples.
Table~\ref{table:train_val_test_split} shows the train-validation-test simulator query breakdown. 
The challenge was to find settings for $m$ and $o$ where the train-val-test proportions for both ex ante and interim are reasonable, and the validation and test sets have higher numbers of observations per $(\vec{\sigma}, \reserve)$ pair to combat noise resulting from different sampled auction configurations (quality scores and valuations for each player). 
It is plausible that---outside of direct comparison with ex ante---interim models achieve slightly lower error than shown in our experiments when the training set has more than $62.5\%$ of simulator queries (training examples).
Also, with fixed $\nsimq$ and more extreme $\nobs$ values (e.g., 1-2 or 1000), deviation payoff errors were high enough that we did not pursue formal hyperparameter tuning for these cases. 

\input{tables/train_val_test_split}

We provide ex ante payoff statistics for all 6 training sets in Table~\ref{table:train_payoff_stats} and  all test sets in Table~\ref{table:test_payoff_stats}.
``ML Test Set'' refers to the noisy test set that uses the same sampling process to approximate the ground truth as used for training. 
Results presented in Fig.~\ref{fig:ml_test_results} are based on ML Test Set \#1; in App.~\ref{app:ml_test} we provide additional results based on ML Test Sets \#2 and \#3. 
``Extra Test Set'' refers to the fine-grained grid dataset that is much larger and less noisy than any of the ML test sets; this dataset is described further in App~\ref{app:extra_test}. 

\input{tables/train_payoff_statistics}

\input{tables/test_payoff_statistics}

\subsubsection{Hyperparameter Tuning}

We conduct hyperparameter tuning over the grid of hyperparameters specified in Table~\ref{table:hyperparameter_grid}.
For each payoff function (ex ante or interim) and each of the six training sets, we explore 14 different architectures. 
These architectures are variations of architectures used by \citet{sokota2019learning,gatchel2023learning} as well as other standard ML architectures.
For each architecture, we draw 50 random hyperparameter configurations uniformly from the listed ranges or values in Table~\ref{table:hyperparameter_grid}, producing $14\cdot 50 = 700$ candidate models per function-dataset pair. 
Note that "r" denotes a ReLU layer, the number indicates the nodes in that layer, "{r128}x6" represents six 128-node ReLU layers, and "{[}r128;r128{]}x2" denotes two consecutive residual blocks with two 128-node ReLU layers per block and an additive skip connection between blocks.
Also, "h" denotes a strategy head, "s" indicates a skip-concat connection from the input layer, and "lin" represents a linear layer.

\input{tables/hyperparameter_grid}

With interim models, the training loss was interim mean-squared error and the validation loss was marginalized interim mean squared error using the 300 specific types associated with the target validation payoff samples for each $(\vec{\sigma}, v)$ pair. 
We explored tuning interim models based on interim or marginalized interim loss and found that models tuned based on marginalized interim loss generalized slightly (but consistently) better based on both metrics. 

\input{tables/model_hyperparameters}

\subsection{ML Test Set Results}
\label{app:ml_test}

\begin{figure*}[ht]
\centering
\begin{tabular}{cc}
\includegraphics[width=.4\columnwidth]{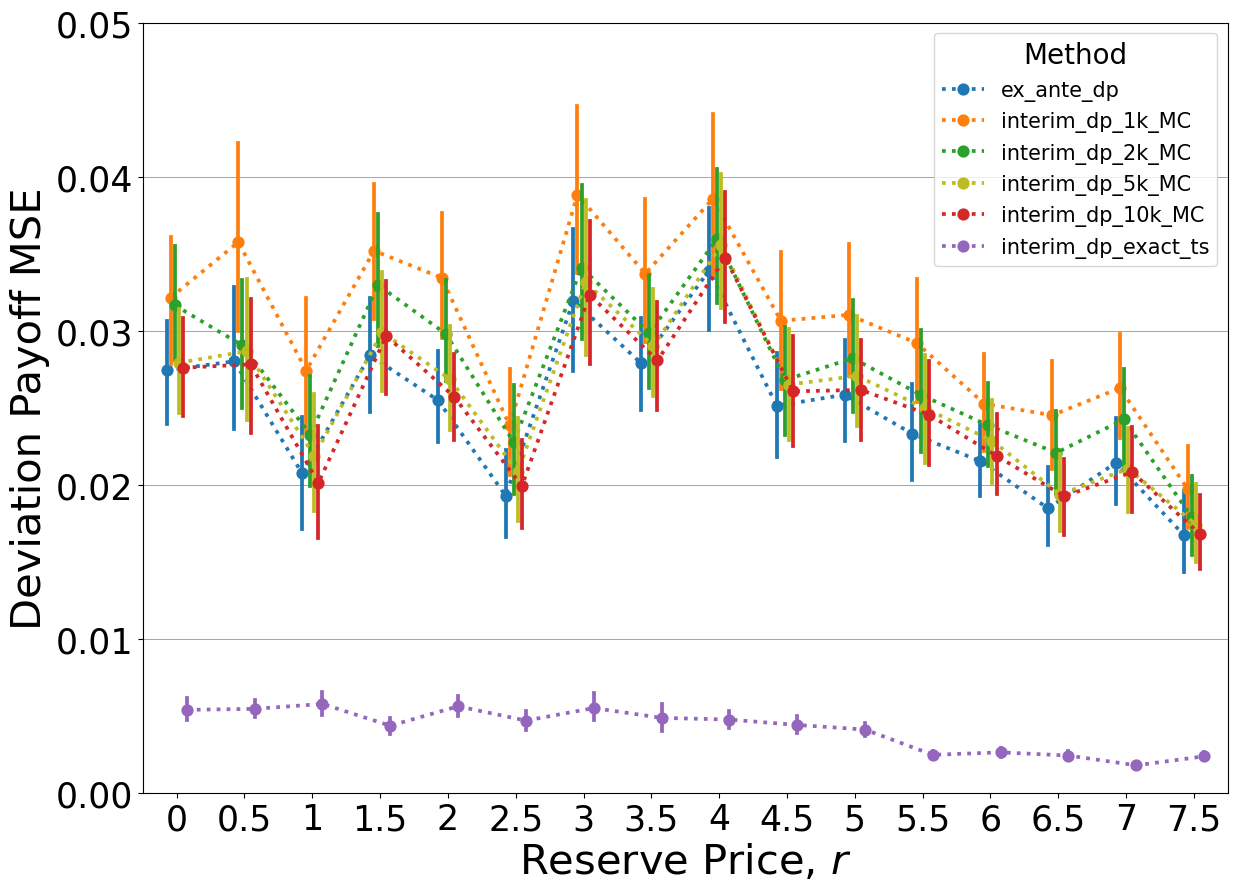} &
\includegraphics[width=.4\columnwidth]{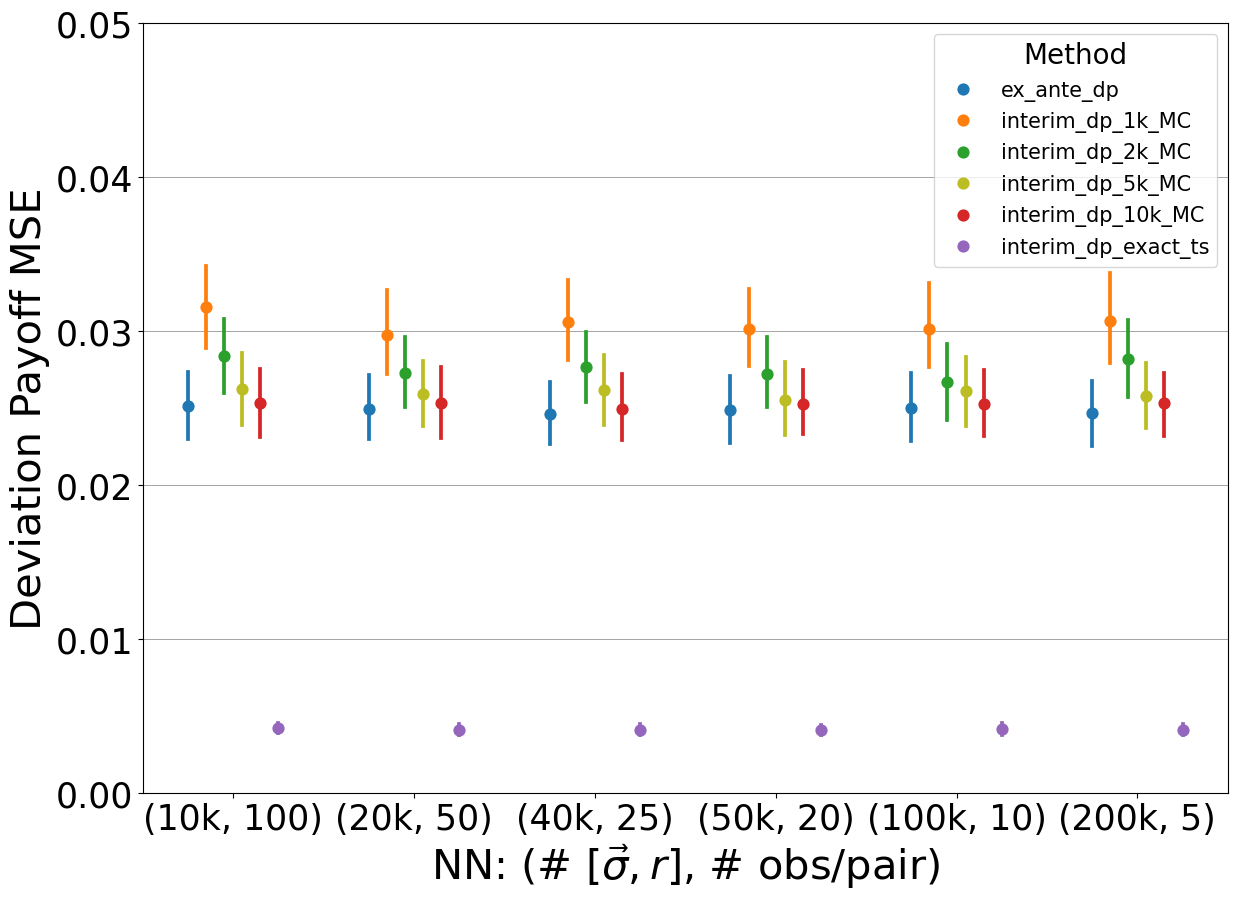} \\
(a) & (b) \\ 
\includegraphics[width=.4\columnwidth]{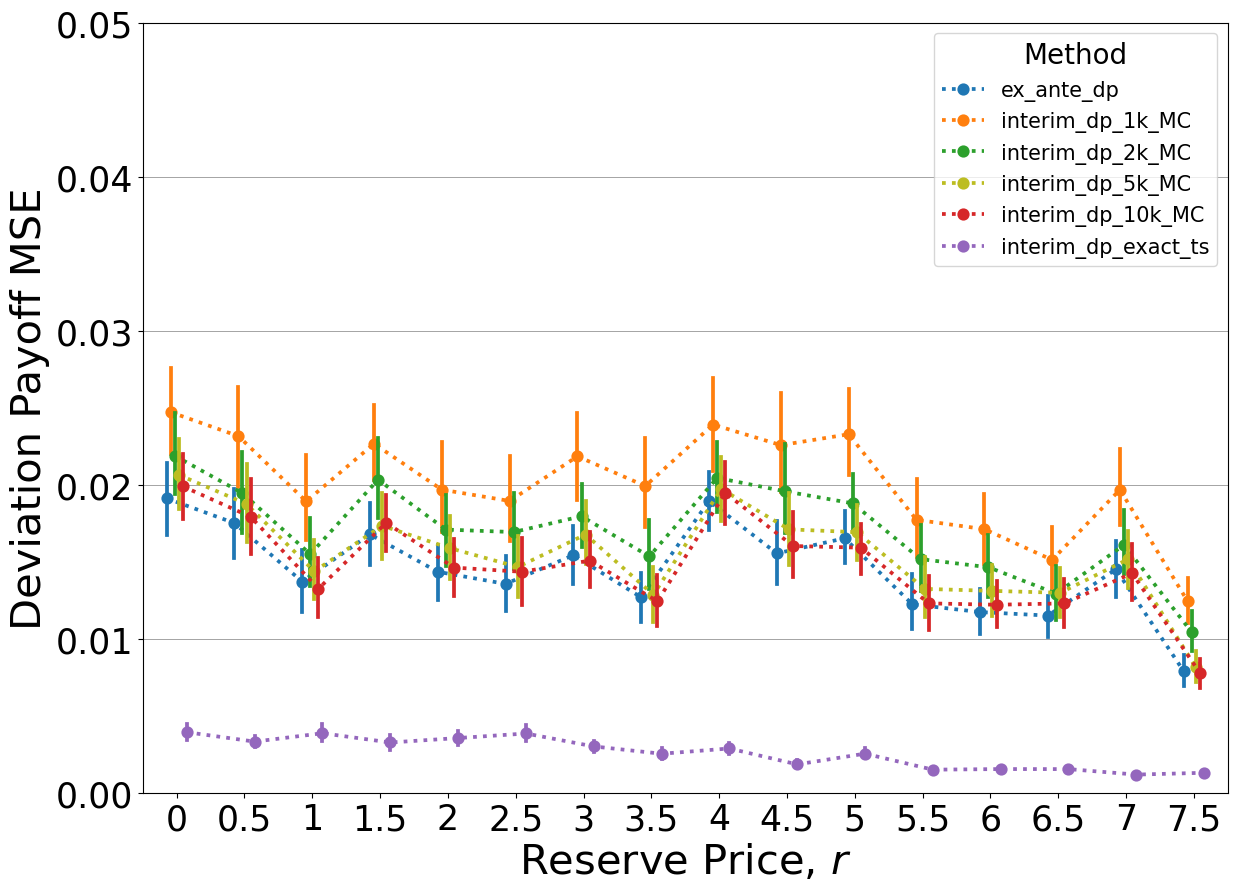} & 
\includegraphics[width=.4\columnwidth]{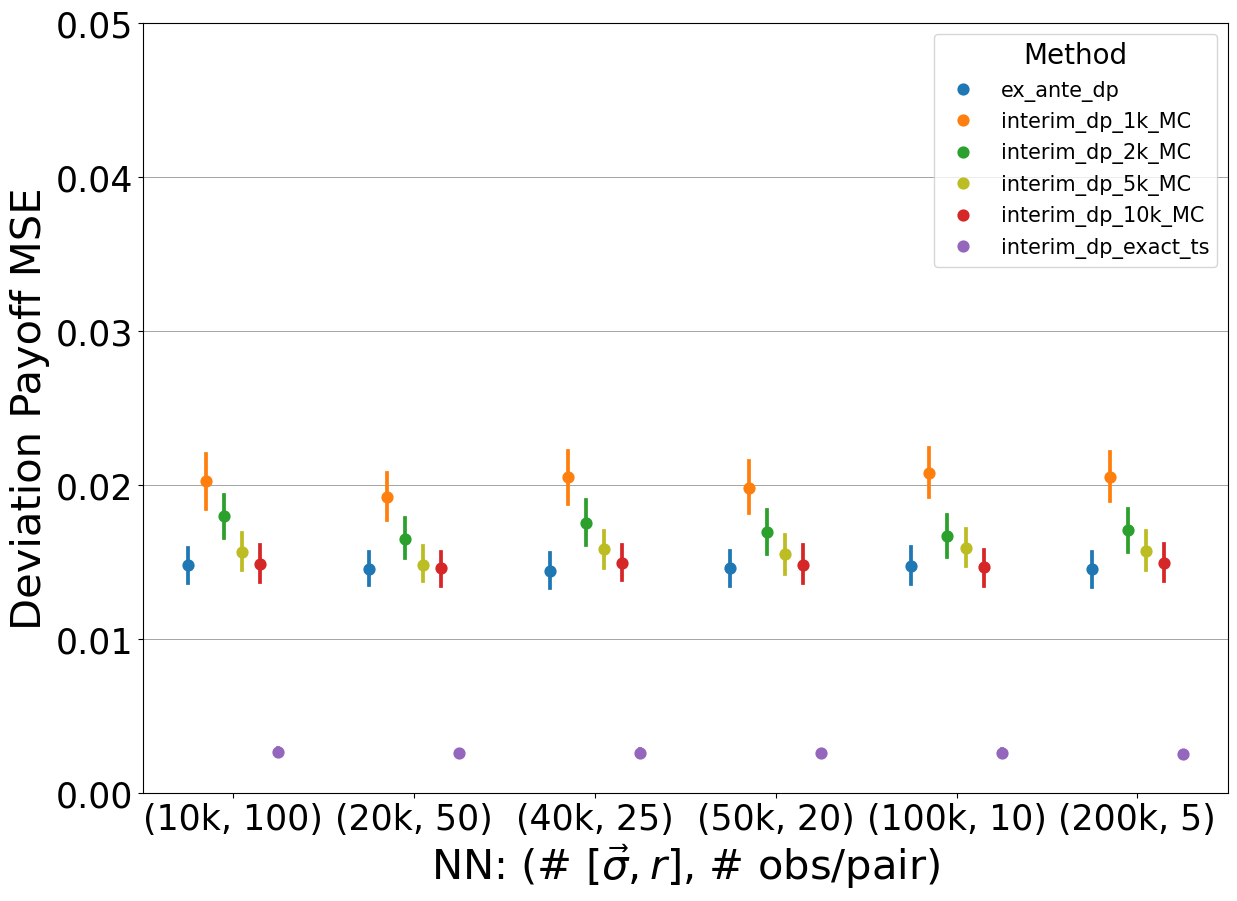} \\
(c) & (d) \\
\includegraphics[width=.4\columnwidth]{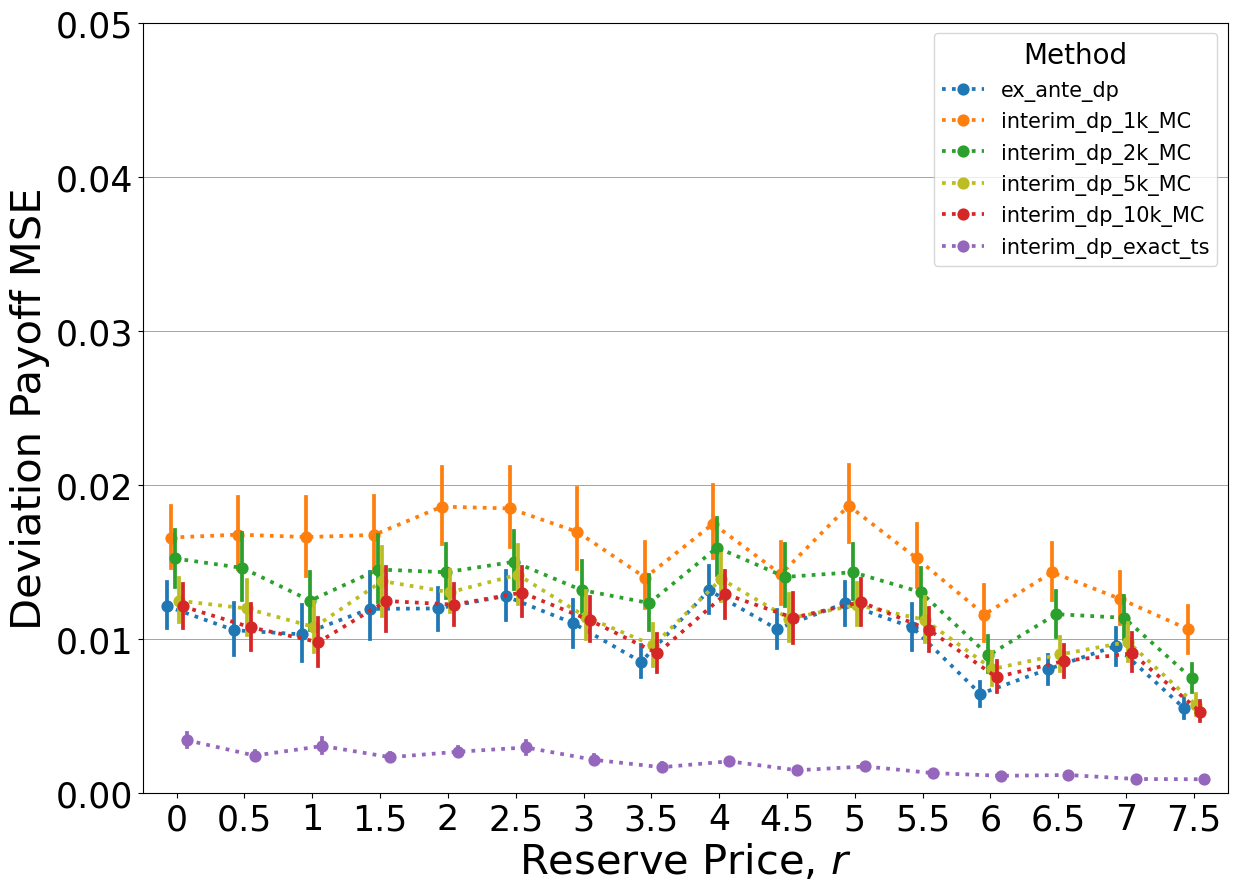} & 
\includegraphics[width=.4\columnwidth]{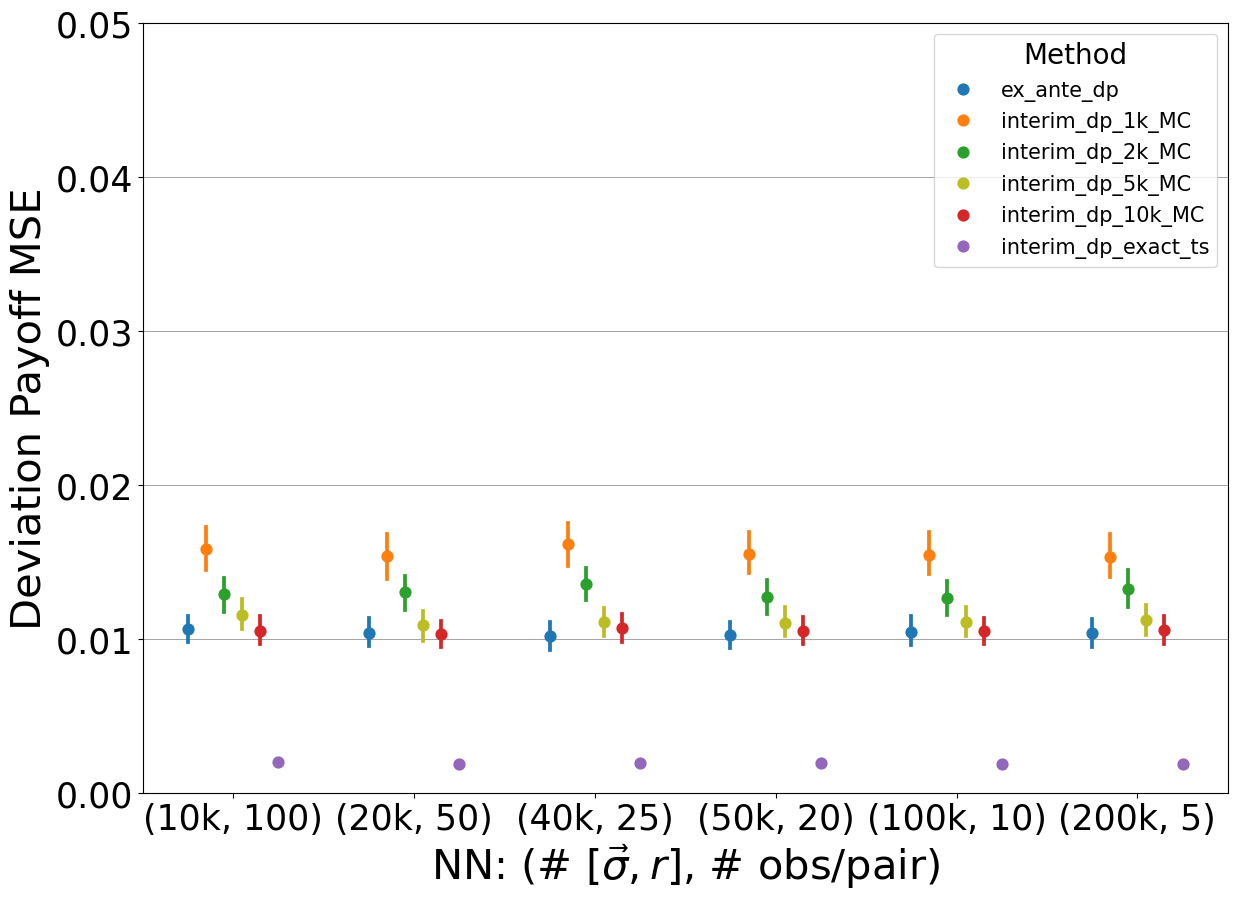} \\
(e) & (f) 

\end{tabular}
\caption{Top: ML Test Set \#1 (300 obs). Middle: ML Test Set \#2 (500 obs). Bottom: ML Test Set \#3 (700 obs).}
\label{fig:ml_test_results_more_obs}
\end{figure*}

ML Test Set \#1 is a realistically sized test set for a large empirical game family, and contains 1000 $(\vec{\sigma}, \reserve)$ pairs and 300 observations per pair. 
Even with the increased number of observations compared to training---the training dataset with the most observations only has 100---this dataset is still noisy. 
Fig.~\ref{fig:ml_test_results_more_obs} shows MSE results for two additional datasets which have the same 1000 $(\vec{\sigma}, \reserve)$ mixtures, and either 500 observations per pair (ML Test Set \#2) or 700 observations per pair (ML Test Set \#3). 
These results confirm that increasing test observations reduces error magnitude. 
We also show marginalized interim MSE when the marginalization is calculated using the specific deviator $q$ and $\val$ values for each $(\vec{\sigma}, \reserve)$ pair in the given test dataset (as opposed to randomly generated $q$ and $\val$). 
While this information would not in practice be available for model inference, it can help contextualize the test results, as it is a more accurate estimate of the interim model's overall error. 

\clearpage 
\newpage
\subsection{Fine-Grained Grid Mixture Results}
\label{app:extra_test}

In this dataset, for each game instance we include all symmetric mixed strategies with support size up to 3 with probabilities a multiple of 0.1. 
For each $(\vec{\sigma}, \reserve)$, we simulate payoffs for \textit{all opponent profiles}, $\vec{s}$, that can be sampled from $\vec{\sigma}$, eliminating one source of noise entirely. 
Specifically, for each $(\vec{s}, \reserve)$ we run 10,000 simulations of the auction, and compute the average deviation payoff for each strategy. 
We define the \term{true deviation payoff vector} for a given $(\vec{\sigma}, \reserve)$ pair as the average deviation payoff across all $\vec{s}$, weighted by their probabilities in~$\vec{\sigma}$.
Technically this is a high-fidelity estimate of the true deviation payoff vector, but for our purposes it suffices to treat it as the true vector. 

\begin{figure}[ht]
\centering
\begin{tabular}{cc}
\includegraphics[width=.45\columnwidth]{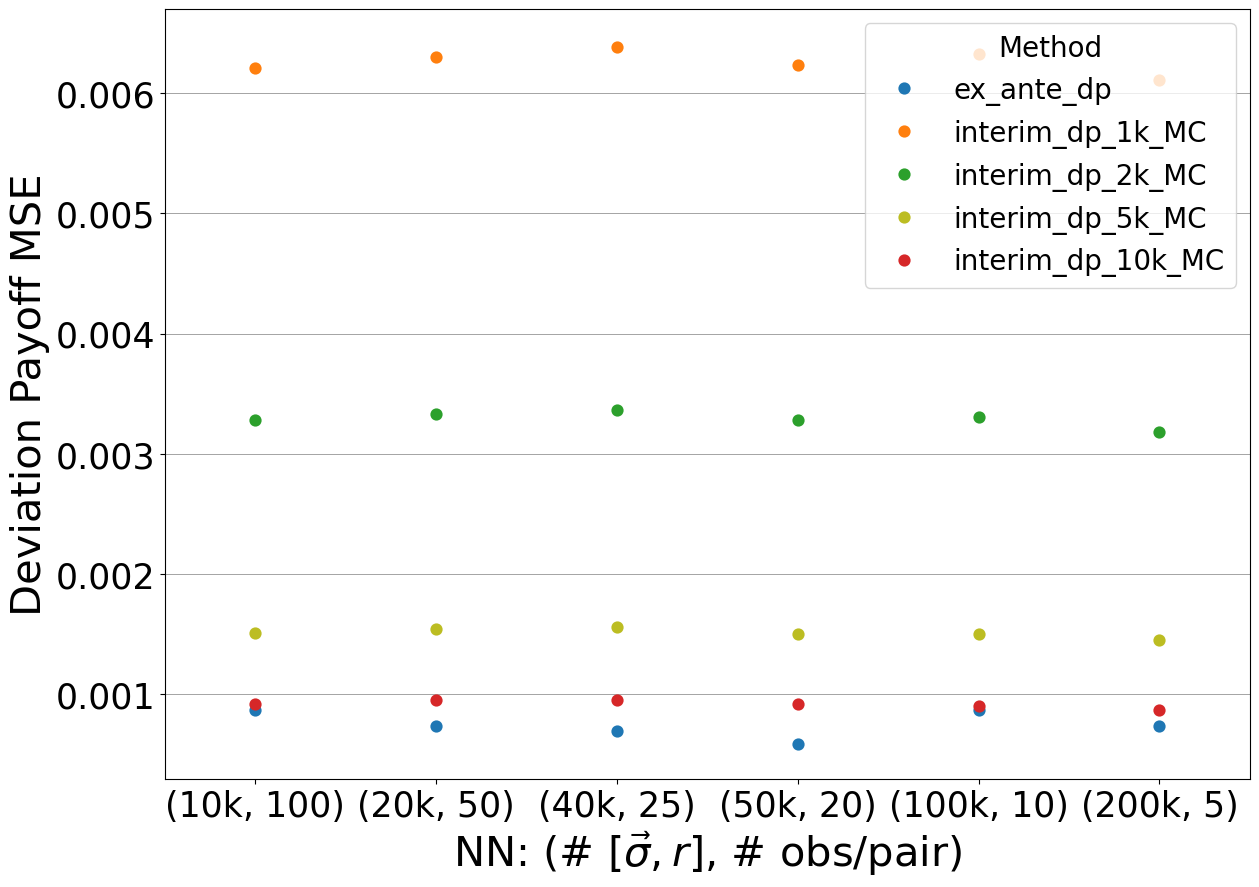} & 
\includegraphics[width=.45\columnwidth]{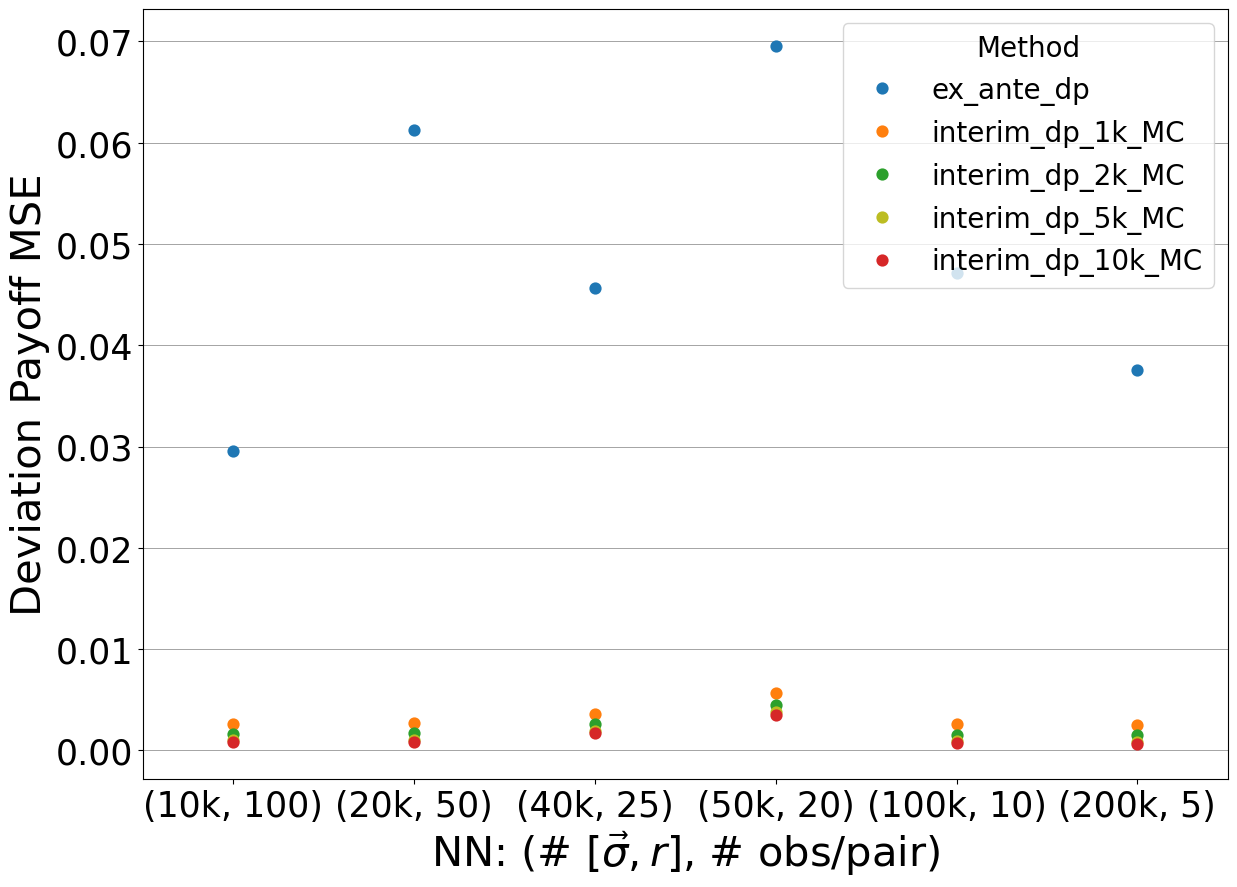} \\
(a) & (b) \\ 
\end{tabular}
\caption{Performance per NN, aggregated over (a) $0.05 \leq r \leq 8$ and (b) $8 < r \leq 15$.}
\label{fig:extra_test_results_per_NN}
\end{figure}

\begin{figure}[ht]
\centering
\begin{tabular}{cc}
\includegraphics[width=.45\columnwidth]{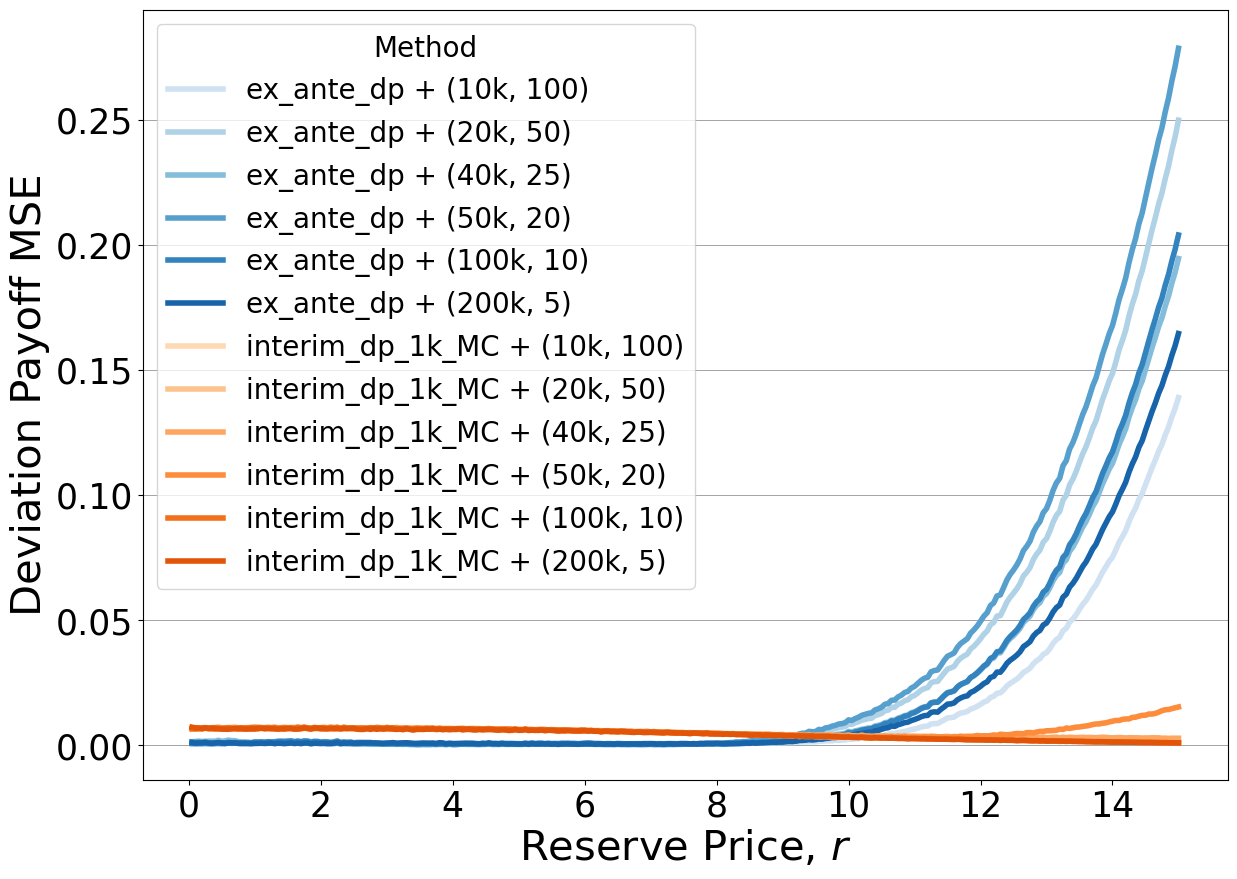} & 
\includegraphics[width=.45\columnwidth]{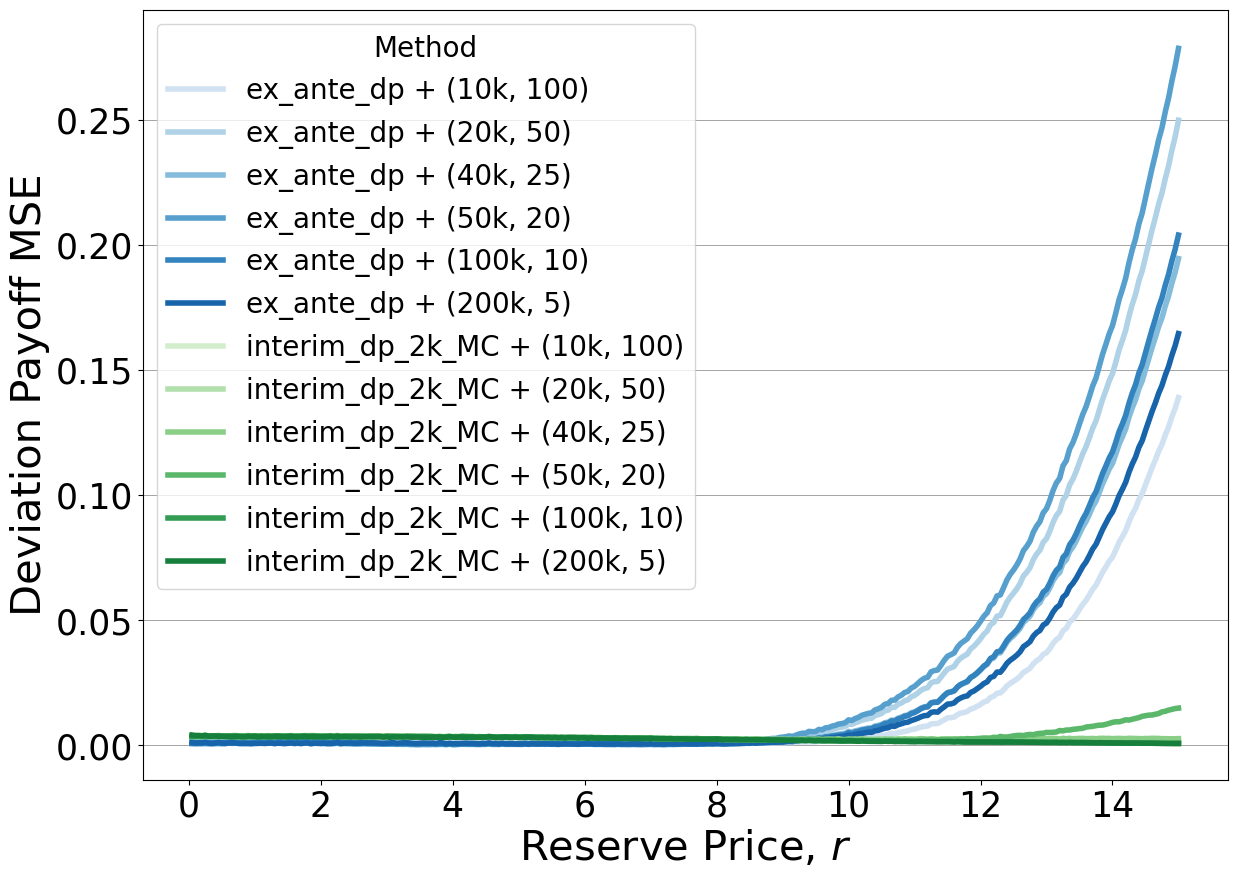} \\ 
(a) & (b)
\end{tabular}
\caption{The same results as Fig.~\ref{fig:extra_test_results_rleq15_orig} except performance is now shown for each learned model for the reserve range $0.05 \leq r \leq 15$.}
\label{fig:extra_test_results_per_NN_1k_fill}
\end{figure}

We present the same results as shown in Figures~\ref{fig:extra_test_results_rleq8}~and~\ref{fig:extra_test_results_rleq15_orig}, but instead show the performance for each of the twelve learned models. 
Fig.~\ref{fig:extra_test_results_per_NN} aggregates over reserve prices, and Fig.~\ref{fig:extra_test_results_per_NN_1k_fill} shows performance for each NN across the reserve space.

\subsubsection{Additional Extrapolation Results}

For each $\maxr \in \{1,\dots,7\}$ and each of the six original training datasets, we construct new datasets by filtering out data points with a reserve setting above $\maxr$.
Table~\ref{tab:extrap_dataset_stats} gives the proportion of training examples from the original dataset that are included in the new dataset with the given maximum reserve. 
Each unit decrease in the maximum reserve reduces the dataset size by 12.5\% relative to the original training dataset. 
The number of observations remains the same as in the corresponding original training dataset; only the number of $(\vec{\sigma}, \reserve)$ pairs in the dataset is reduced. 

\begin{table}[ht]
\caption{Proportion of training examples retained from the original training dataset for each maximum reserve value.}
\begin{tabular}{c|c|c|c|c|c|c|c|c}
$(\vec{\sigma}, \reserve)$ Pairs & Obs/Pair & $r \leq 1$ & $r \leq 2$ & $r \leq 3$ & $r \leq 4$ & $r \leq 5$ & $r \leq 6$ & $r \leq 7$ \\
\hline
10000  & 100 & 0.119 & 0.243 & 0.375 & 0.494 & 0.622 & 0.749 & 0.876 \\
20000  & 50  & 0.123 & 0.248 & 0.372 & 0.495 & 0.623 & 0.751 & 0.877 \\
40000  & 25  & 0.121 & 0.245 & 0.371 & 0.494 & 0.620 & 0.748 & 0.874 \\
50000  & 20  & 0.121 & 0.247 & 0.371 & 0.495 & 0.620 & 0.748 & 0.875 \\
100000 & 10  & 0.121 & 0.245 & 0.372 & 0.496 & 0.621 & 0.748 & 0.874 \\
200000 & 5   & 0.123 & 0.247 & 0.374 & 0.499 & 0.624 & 0.749 & 0.875 \\
\hline 
\multicolumn{2}{r|}{\textbf{Average}}& \textbf{0.121} & \textbf{0.246} & \textbf{0.373} & \textbf{0.496} & \textbf{0.622} & \textbf{0.749} & \textbf{0.875}                       
\end{tabular}
\label{tab:extrap_dataset_stats}
\end{table}

\begin{figure}[ht]
\centering 
\begin{tabular}{cc}
    \includegraphics[width=.45\columnwidth]{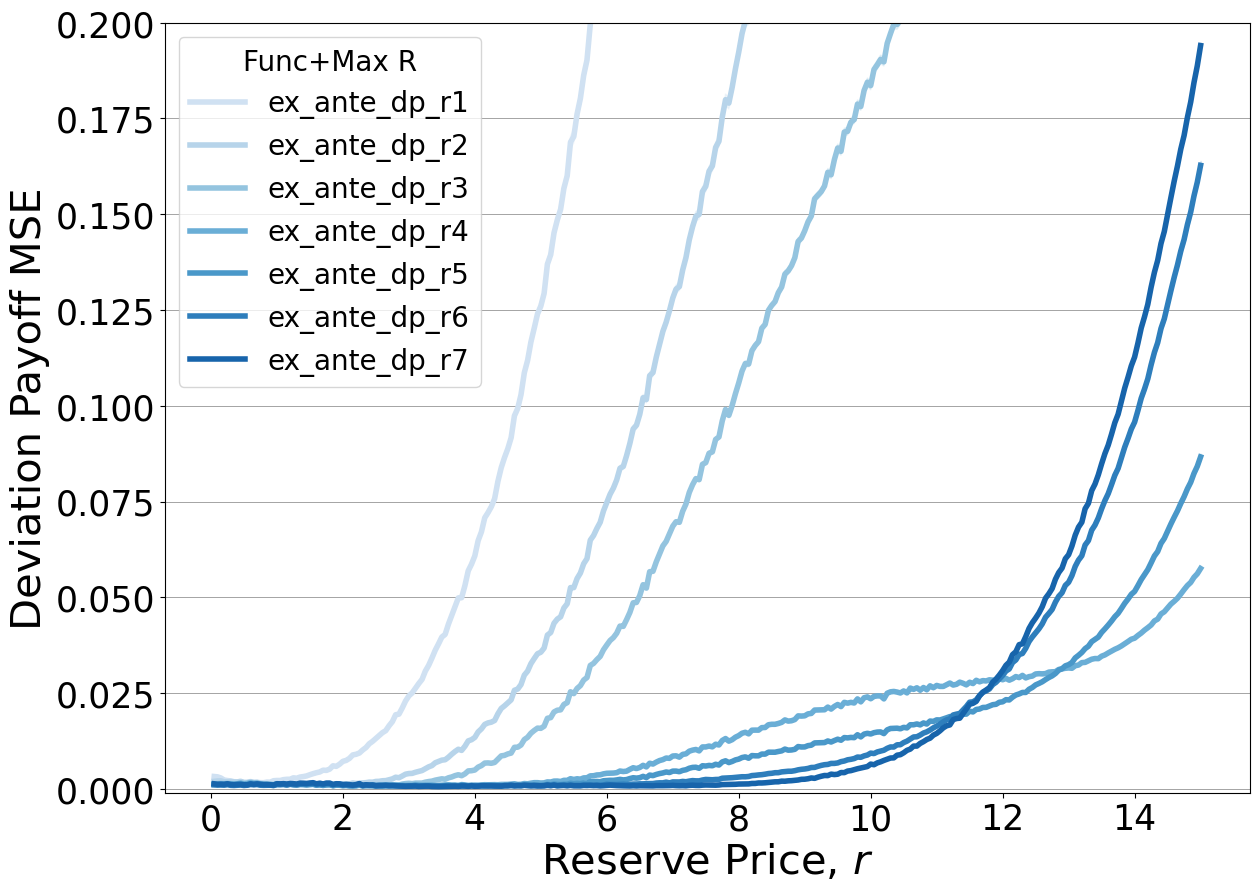} & \includegraphics[width=.45\columnwidth]{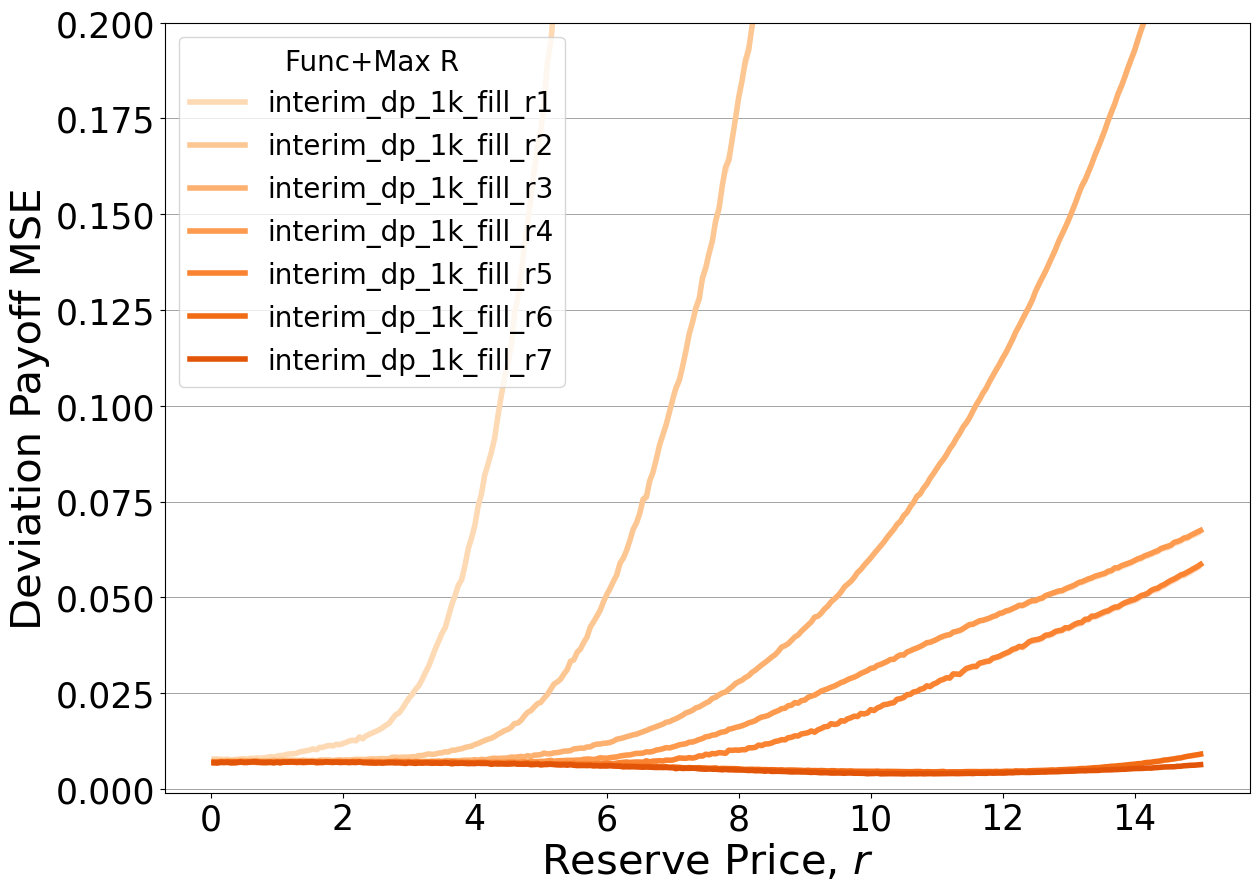} \\
    (a) & (b)
\end{tabular}
\caption{(a) Ex ante models have a non-monotonic extrapolation trend with increasing $\maxr$, possibly due to reduced payoff smoothness at higher reserves. (b) Interim model extrapolation improves consistently as $\maxr$ increases.}
\label{fig:extra_extrapolation_results_all_max_r}
\end{figure}

Fig.~\ref{fig:extra_extrapolation_results_all_max_r} shows (a) ex ante and (b) interim deviation payoff errors for models trained on different ranges of reserve settings.
For $\maxr = 1$ to $\maxr = 3$, both ex ante and interim models achieve low deviation payoff error on the respective trained range, but the error quickly exceeds $\varepsilon=0.01$ in extrapolation. 
This trend is to be expected because each neural network is trained on only approximately 12.5\%--37.5\% of the original dataset.
The interim model is likely more affected by smaller datasets because, as the dataset shrinks, a larger fraction of simulator queries are used for repeated observations of the same $(\vec{\sigma}, \reserve)$ profile. 
If we were to generate smaller training datasets from scratch using the same total number of simulator queries, we would likely select an $\nmix$-to-$\nobs$ ratio closer to that of the original six datasets.

As $\maxr$ increases, interim extrapolation performance steadily improves.
Ex ante extrapolation performance improves up to $\maxr = 4$, and then degrades for higher maximum reserves.
A plausible explanation for this difference is that the smoothness of the payoff function varies across the parameter space. 
At higher reserves, the payoff function becomes less smooth: only a fraction of bidders can meet the reserve, but those who do receive large payoffs. 
The interim model, by conditioning on type, can learn these sharp transitions that depend heavily on deviator type, leading to improved extrapolation performance as $\maxr$ increases. 
In contrast, the ex ante model---trained on payoffs in expectation over types---has less information and may overfit to the abrupt patterns present in the training data for high reserves. 
Consequently, an ex ante model trained on a moderate reserve range (e.g., $\reserve \leq 4$) learns a smooth mapping that extrapolates relatively well, whereas an ex ante model trained on a larger range (up to $\reserve = 8)$ overfits to high-reserve irregularities that degrade its ability to generalize beyond the trained range. 
Such a model still performs well on unseen mixed-strategy profiles within the trained range, as shown in Fig.~\ref{fig:extra_extrapolation_results_all_max_r}, but fails to extrapolate reliably beyond it. 
In conclusion, when trained on ample data across a sufficiently wide parameter range, interim---but not ex ante---models exhibit robust and consistent extrapolation.

\clearpage


\section{Using the Learned Game-Family Model}
\label{app:use_learned_model}

\subsection{Deriving Equilibria}
\label{app:nash_approximation}

We run replicator dynamics from 11 full-support mixed-strategy profiles: the uniform random mixture and, for each strategy $s_j$ of the 10 strategies, a mixed strategy with probability $\frac{2}{11}$ on strategy $s_j$ and $\frac{1}{11}$ on the others. 
For interim models, we separately run RD with 1000 and 2000 Monte Carlo type samples for marginalization. 
For each returned $\vec{\sigma}$, we use $\hat{u}$ to predict regret.

The \term{support of a candidate $\varepsilon$-BNE} is the set of strategies played with at least 0.01 probability. 
Probabilities below 0.01 are set to 0 and the mixture is renormalized. 
We compute the true deviation payoff vector as in \S\ref{sec:extra_test_set} (described further in App.~\ref{app:extra_test}), simulating 10,000 auctions for each $\vec{s}$ that could be sampled from the (possibly renormalized) candidate mixture.
Using this true deviation payoff vector we can calculate \term{true regret}, and regard a candidate $\varepsilon$-BNE as confirmed if true regret is at most $\varepsilon=0.01$. 

Tables~\ref{table:mix_class_rleq8}-\ref{table:mix_class_rgt8} show classification results for for $\reserve \leq 8$ and $\reserve > 8$, respectively.

\input{tables/mix_class_rleq8}

On the trained range, all ex ante and interim models reject less than 3\% of mixtures.
In extrapolation, all interim models reject less than 5\% of mixtures, but ex ante models reject over 40\% of mixtures on average. 
Additionally, for certain mixed strategies, the deviation payoff vector predicted by an ex ante model is all zeros. 
This disrupts RD, as multiplying by a zero vector "kills" the iteratively improving mixed strategy, creating what we call a \term{dead mixture}. 
On average, over 8\% of ex ante mixtures were classified as dead mixtures.
Overall, ex ante models confirm less than 50\% of mixtures as equilibria, contributing to the significant holes in the expected revenue curves shown in Fig.~\ref{fig:emd_grid_optimization}.

\input{tables/mix_class_rgt8}

\begin{figure}
    \centering
    \includegraphics[width=.5\columnwidth]{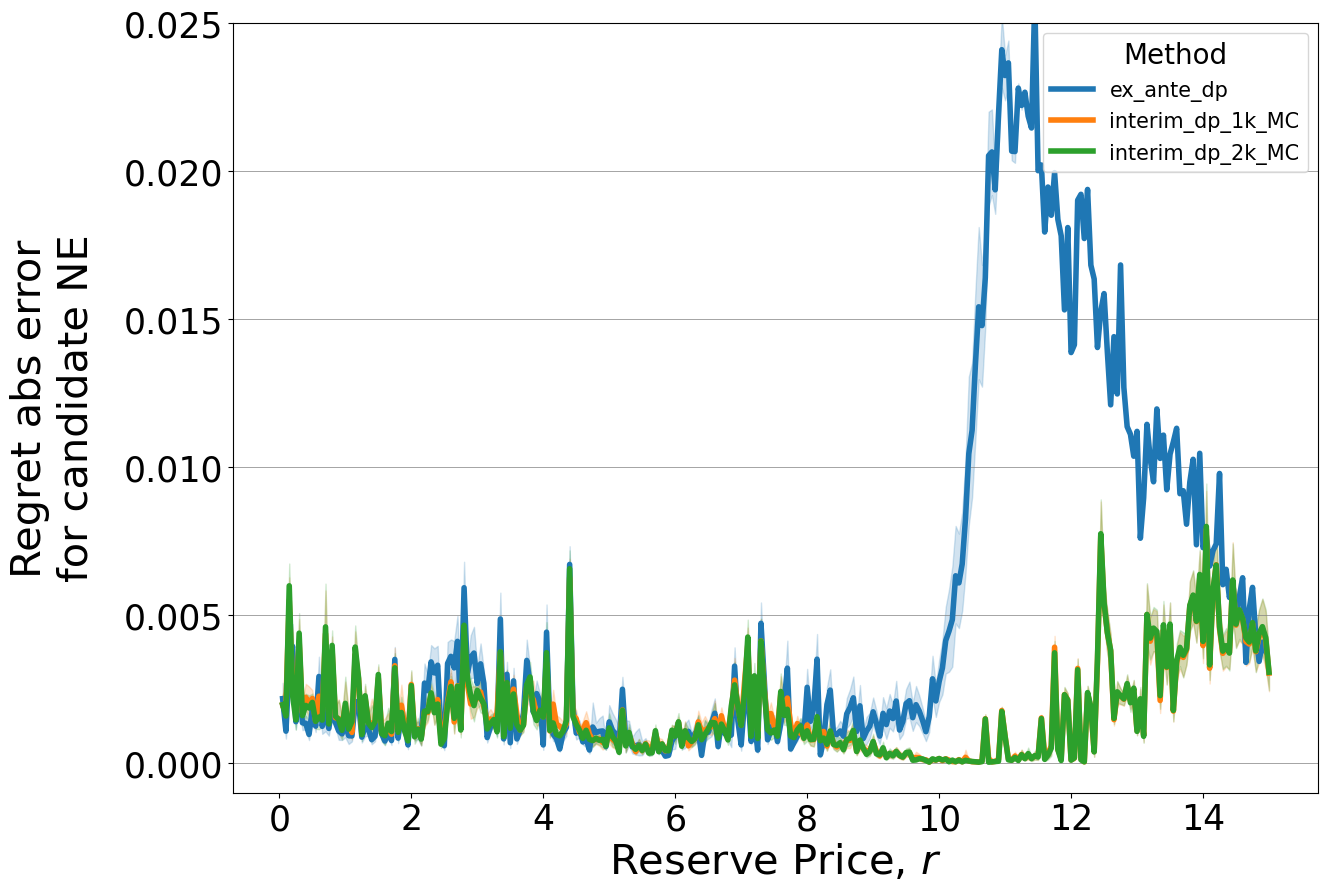}
    \caption{Ex ante and interim regret error is similar on the trained range, and ex ante regret error is larger in extrapolation.}
    \label{fig:cand_ne_regr_abs_err_over_r}
\end{figure}

Fig.~\ref{fig:cand_ne_regr_abs_err_over_r} shows mean absolute error between predicted and true regret for ex ante and interim methods, with 95\% confidence intervals shaded.
Each point represents the average regret absolute error across all candidate $\varepsilon$-BNE returned by RD using each of the 6 trained models.%
\footnote{For consistent comparison, both predicted and true regret are computed on the truncated and renormalized candidates.}

\begin{figure*}[ht]
\centering
\begin{tabular}{ccc}
\includegraphics[width=.47\columnwidth]{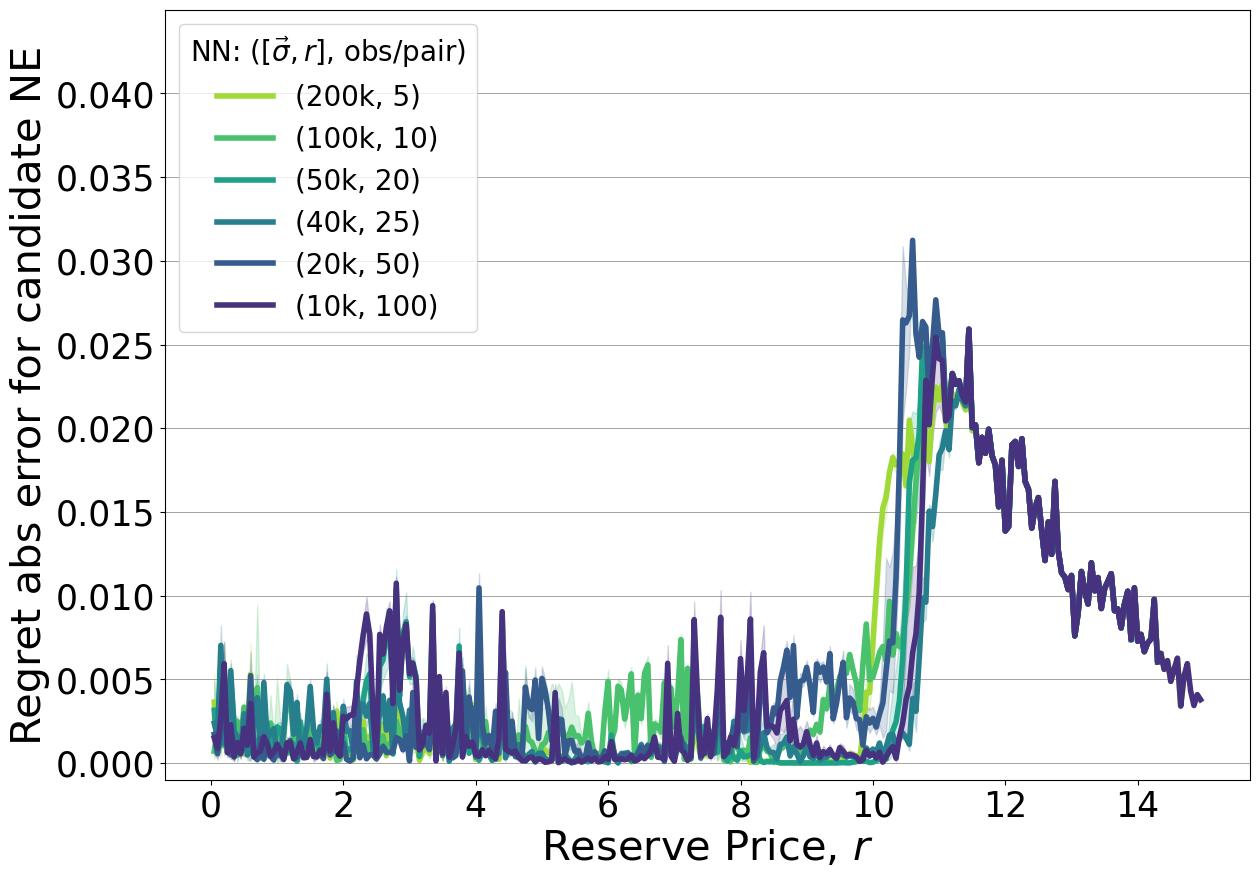} && 
\includegraphics[width=.47\columnwidth]{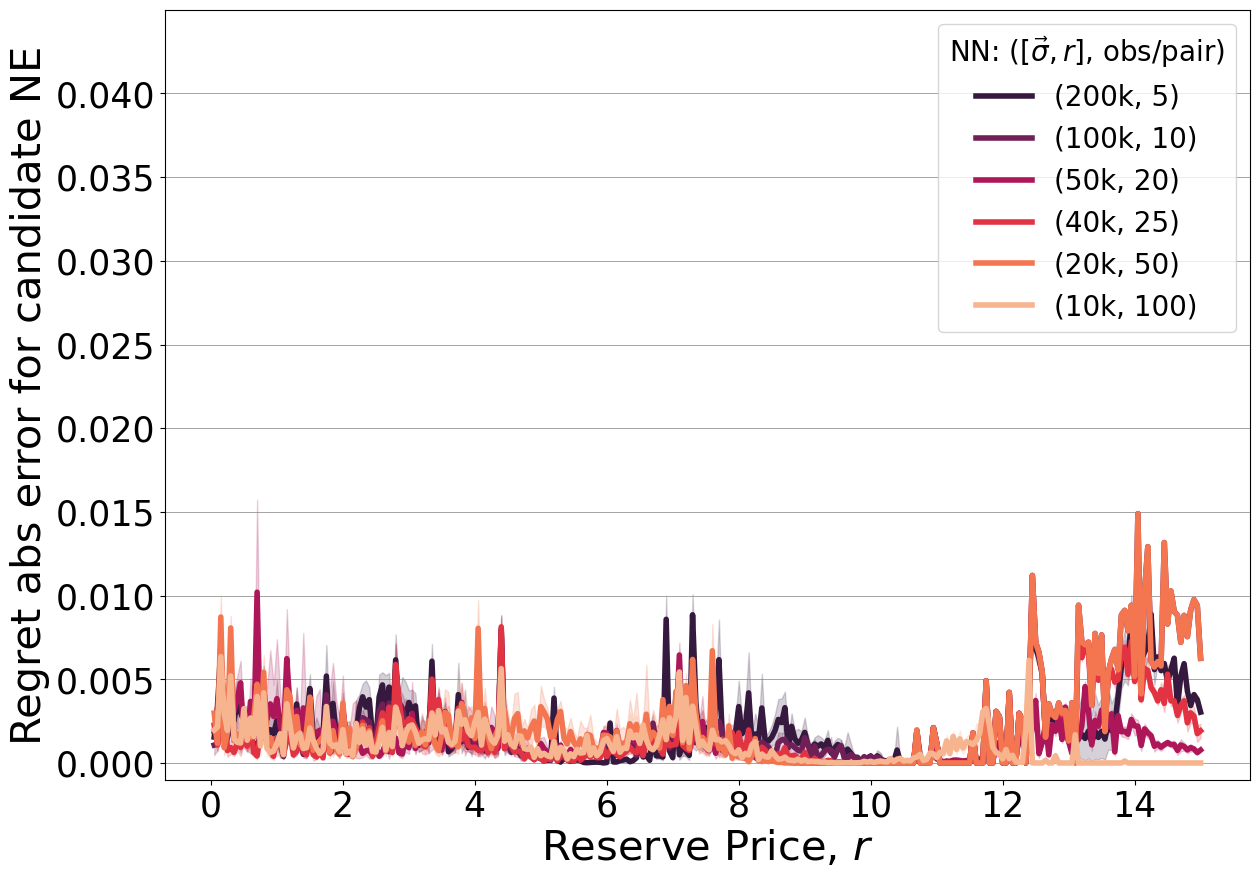} \\
(a) && (b) \\ 
\end{tabular}
\caption{Regret absolute error for candidate equilibria found using each (a) ex ante model and (b) interim model with 1000 marginalization type samples.}
\label{fig:ea_vs_ei_cand_ne_regr_abs_err}
\end{figure*}

Fig.~\ref{fig:ea_vs_ei_cand_ne_regr_abs_err} shows the same results as in Fig.~\ref{fig:cand_ne_regr_abs_err_over_r} except performance is now shown for (a) each ex ante model and (b) each interim model.


\subsection{EMD}
\label{app:emd}

\subsubsection{Grid Optimization: Average Revenue}
\begin{figure}
    \centering 
    \includegraphics[width=.5\columnwidth]{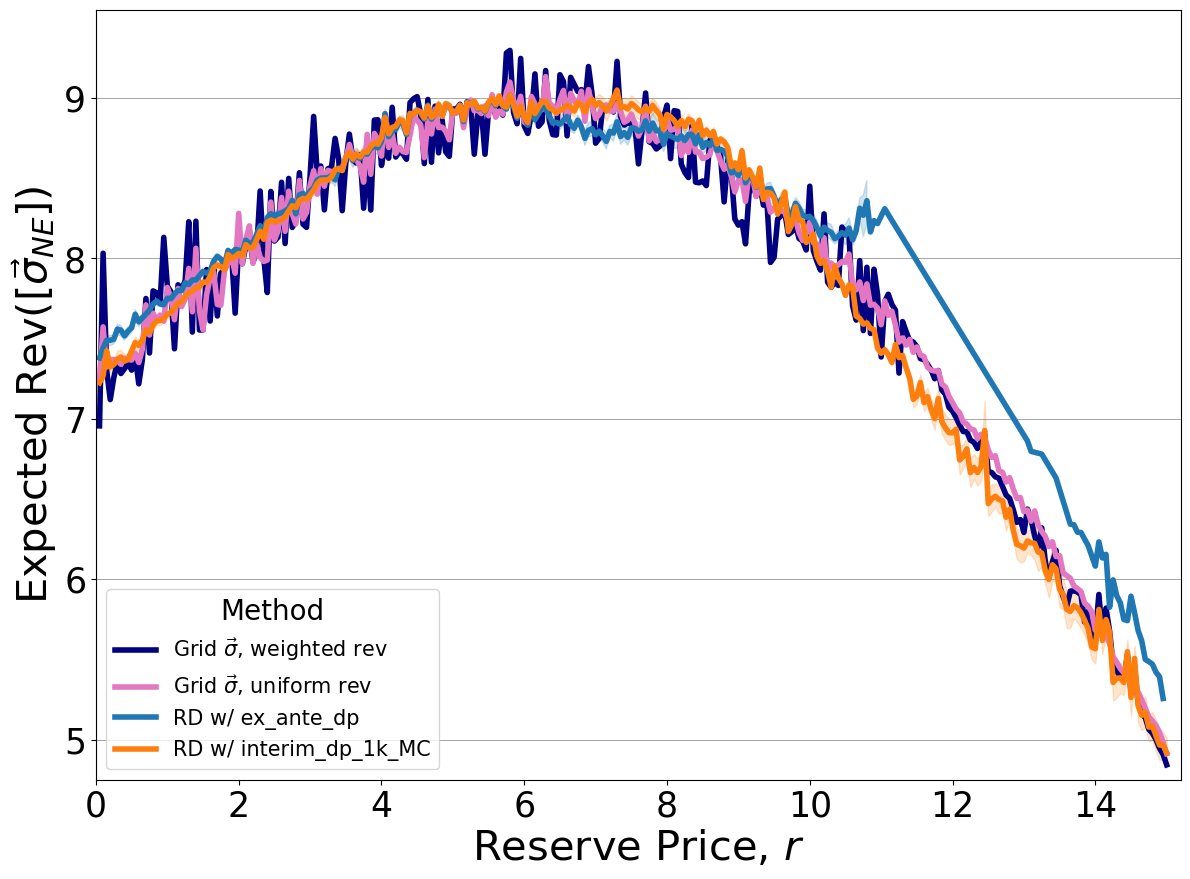}
    \caption{Four methods to aggregate expected revenue in equilibrium produce similar curves.}
    \label{fig:emd_grid_optimization_overall}
\end{figure}

Using a grid of all 3-support\footnote{Most confirmed $\varepsilon$-BNE found by RD have a support size of at most 3.} mixed-strategy distributions across 10 strategies with increments of 0.01 and for all 300 game instances, we calculate true regret, identify $\varepsilon$-BNE, and compute true expected revenue.
For each game instance, the expected revenue is the uniform average revenue across all grid-mixture $\varepsilon$-BNE. 
However, this uniform average may overemphasize higher-support equilibria, so we also compute weighted revenue: for each possible 3-strategy subset $Y \subset S$, we identify all equilibria whose support is contained within $Y$, and compute the average revenue for these equilibria. 
The weighted average revenue is the average revenue across all~$Y$.
Fig.~\ref{fig:emd_grid_optimization_overall} shows ex ante and interim expected revenue curves, aggregated over six models, as well as expected revenue for grid equilibria.
Ex ante and interim expected revenue curves resemble those based on grid equilibria, supporting the practical effectiveness of the learning approaches for parameter optimization.

\subsubsection{Grid Optimization: Worst-Case Revenue}
\begin{figure*}[ht] 
\centering
\begin{tabular}{cc}
\includegraphics[width=.48\columnwidth]{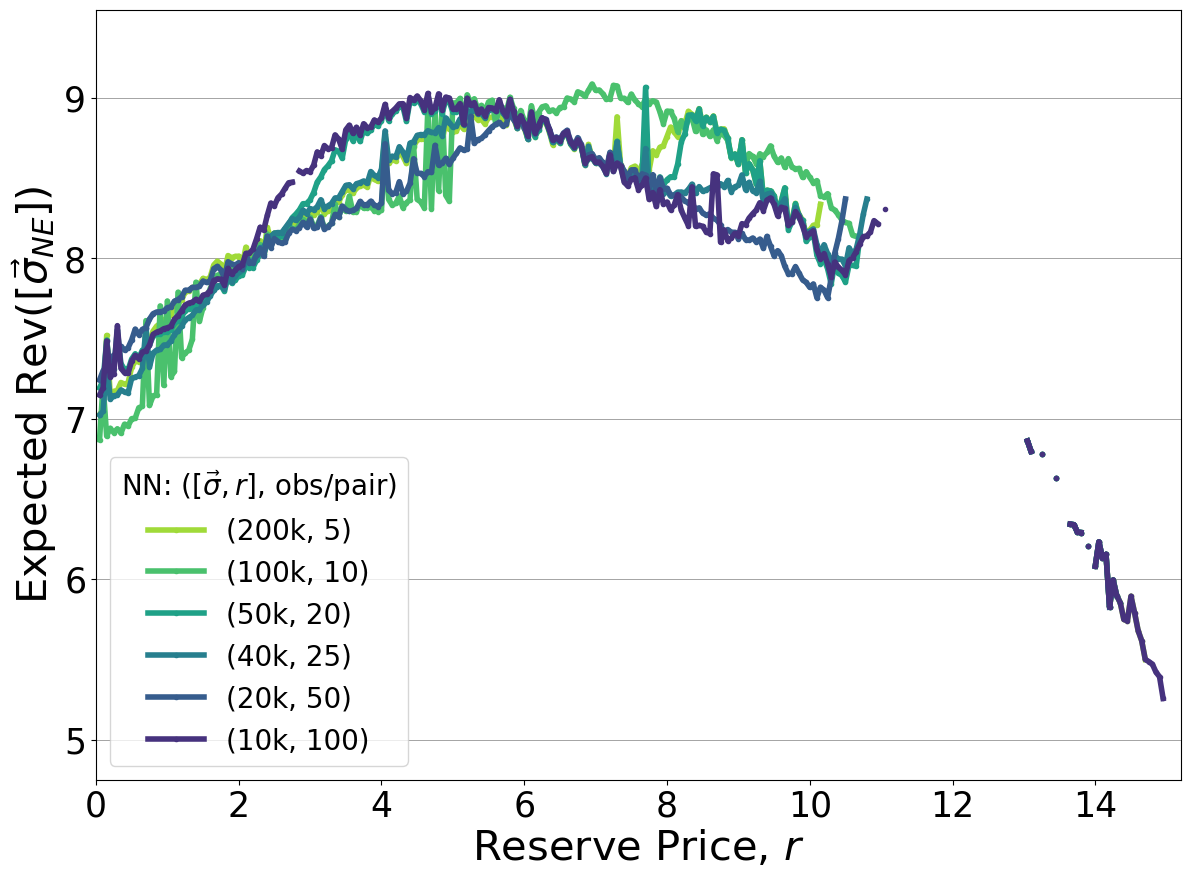} & 
\includegraphics[width=.48\columnwidth]{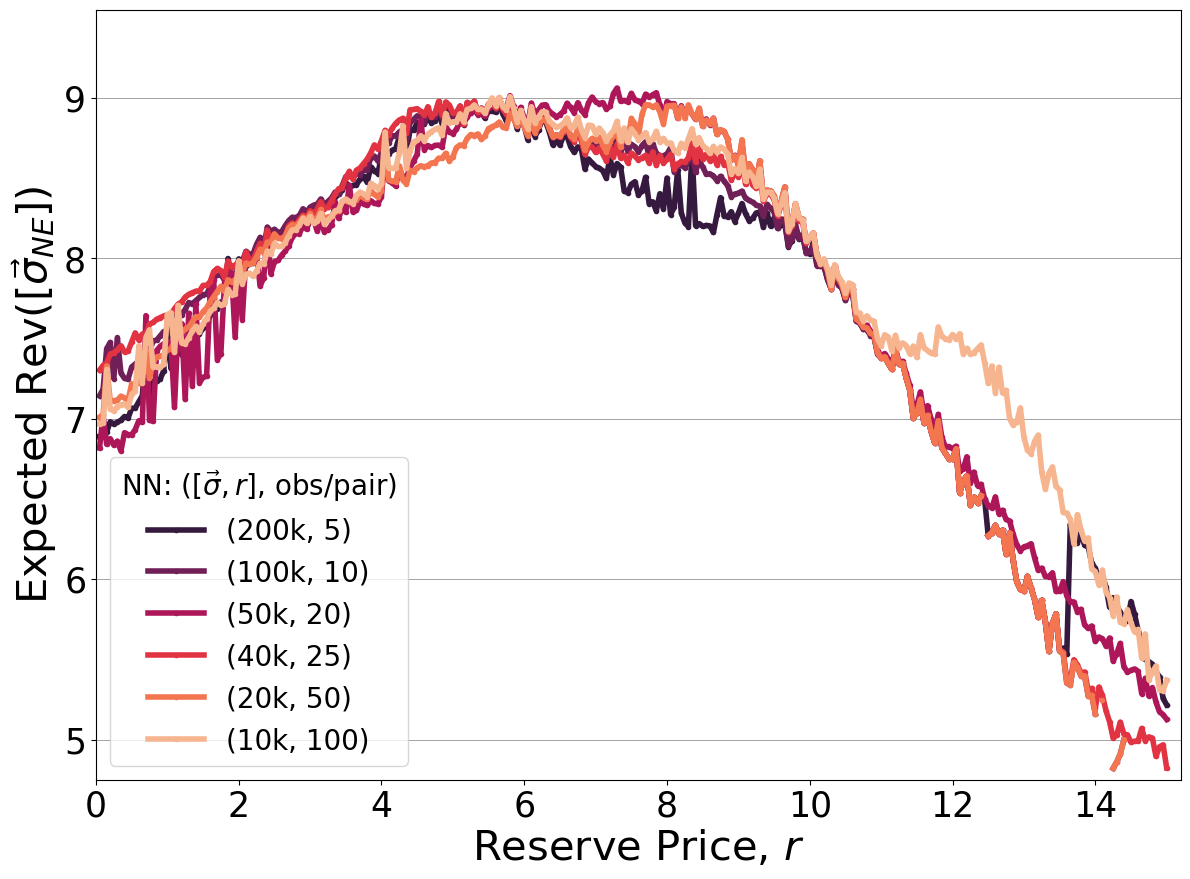} \\ 
(a)  & (b)  \\ 
\end{tabular}
\caption{Worst-case expected revenue for equilibria derived by (a) ex ante and (b) interim models.}
\label{fig:emd_worstcase_grid_optimization}
\end{figure*}

Because we derived sets of $\varepsilon$-BNE for 300 game instances (as described in \S\ref{sec:find_appx_ne}), we can easily perform grid searches using alternative equilibrium-selection methods. 
Fig.~\ref{fig:emd_worstcase_grid_optimization} shows the worst-case expected revenue for equilibria derived by each ex ante and interim neural network. 
The overall shape of the revenue curves resembles that of Fig.~\ref{fig:emd_grid_optimization}, but they are shifted downward, as expected; few models identify equilibria with worst-case expected revenue above 9. 
These revenue curves are also less smooth than those in Fig.~\ref{fig:emd_grid_optimization}, reflecting the sensitivity of this selection method to the particular equilibria derived for each game instance.
The main takeaway is that conducting mechanism design with a chosen equilibrium-selection method becomes more feasible with game-family learning, since it enables efficient derivation of approximate equilibria across a large number of game instances.

\subsubsection{Local Search Optimization}

\begin{figure}[ht]
    \centering
    \includegraphics[width=.5\columnwidth]{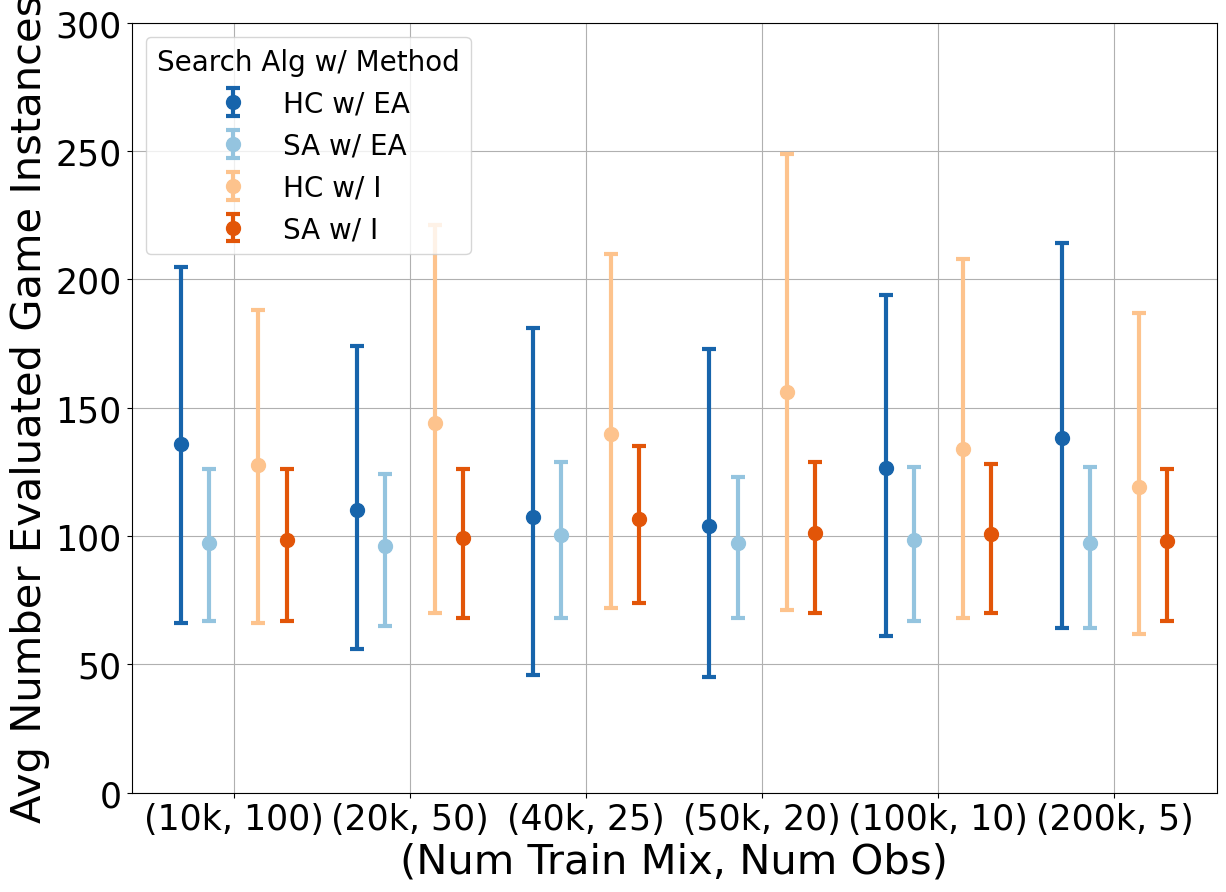}
    \caption{The average number of distinct game instances evaluated in local search experiments with 5 restarts. 
    Bootstrap confidence intervals are computed concurrently with the intervals in Fig.~\ref{fig:local_search_results}. 
    On average, local search with 5 restarts evaluates roughly 100-150 distinct game instances compared to 300 game instances for grid optimization (30-50\%).}
    \label{fig:num_game_instances_evaluated}
\end{figure}

Concurrent with computing bootstrap confidence intervals for optimal expected revenue in equilibrium (Fig.~\ref{fig:local_search_results}), we also compute bootstrap confidence intervals for the average number of game instances explored using local search. 
Specifically, for each of the 100,000 5-restart experiment samples, we determine the number of distinct game instances evaluated across the 5 restarts.
In Fig.~\ref{fig:num_game_instances_evaluated}, we plot the resulting 95\% confidence intervals around the average total number of game instances explored. 

\newpage
\subsection{Computing Piecewise Best Responses}
\label{app:compute_pwbr}

For computing piecewise best responses, we reduce the two type parameters, $q$ and $\val$, to one dimension by taking the product $q\cdot \val$, representing a player's maximum effective bid. 
We partition this single-dimensional type space into 5 roughly equiprobable intervals, shown in Table~\ref{tab:type_intervals}.

\begin{table}[ht]
    \centering
    \begin{tabular}{c|c|c}
        & \textbf{Start} & \textbf{End} \\
        \hline
        $\interval_0$ & 0.        & 1.251829  \\
        $\interval_1$ & 1.251829  & 3.308699  \\
        $\interval_2$ & 3.308699  & 6.312033  \\
        $\interval_3$ & 6.312033  & 10.963049 \\
        $\interval_4$ & 10.963049 & 25.       \\
    \end{tabular}
    \caption{Type partition used for computing piecewise best responses, where type is $q \cdot \val$.  The first column shows the starting cutoff, and the second column shows the ending cutoff.}
    \label{tab:type_intervals}
\end{table}

\begin{figure}
    \centering
    \includegraphics[width=.5\columnwidth]{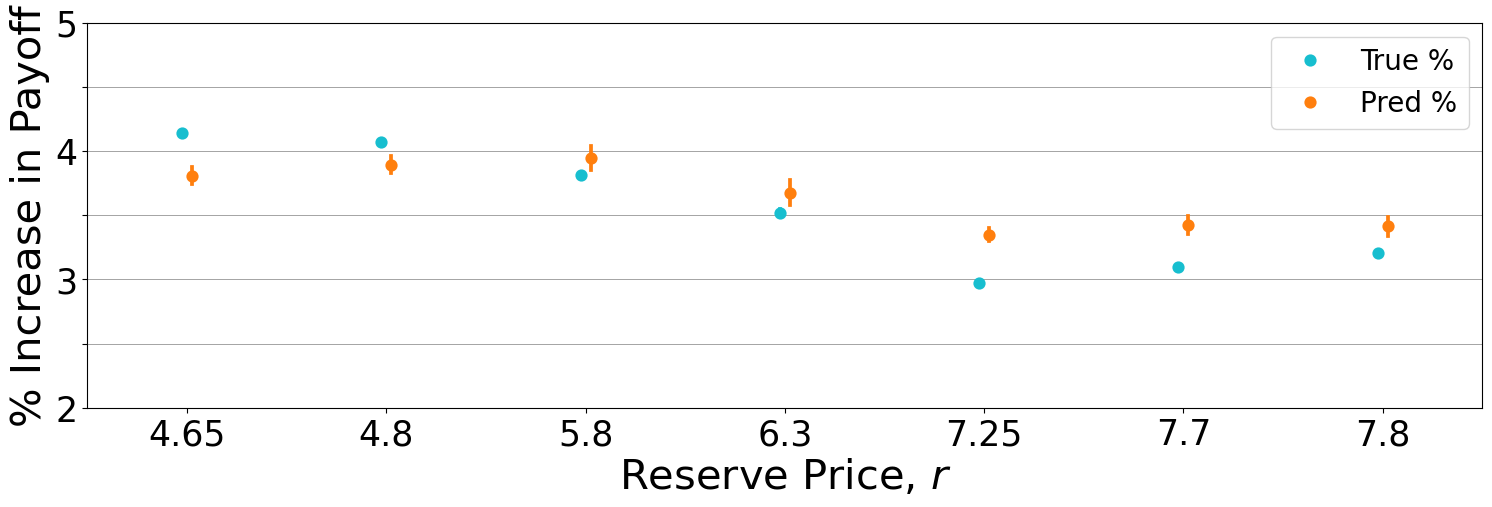}
    \caption{On average, players can gain approximately 3\% payoff or more by playing the piecewise best-response strategy to $\vec{\sigma}_{\BNE}$ compared to playing the $\vec{\sigma}_{\BNE}$.}
    \label{fig:piecewise_pct_incr}
\end{figure}

Recall that Fig.~\ref{fig:piecewise_gain} showed the gain from playing the piecewise strategy compared to playing the $\varepsilon$-BNE while opponents play the $\varepsilon$-BNE. 
Fig.~\ref{fig:piecewise_pct_incr} shows the same results, but now the deviation payoff gain is expressed as the percent increase from playing the equilibrium to playing the piecewise strategy. 

%% file: tables/train_val_test_split.tex
\begin{table*}[ht]
\centering
\begin{tabular}{c|c|c|c|c|c}
Dataset   & $(\sigma, r)$ Pairs & EA Proportion & Obs/pair & Total Queries & Interim Proportion \\ \hline
Train \#1 & 10000                              & 0.8333                          & 100                             & 1000000                             & 0.625                          \\
Val       & 1000                               & 0.0833                          & 300                             & 300000                              & 0.1875                         \\
Test      & 1000                               & 0.0833                          & 300                             & 300000                              & 0.1875                         \\
\hline 
Train \#2 & 20000                              & 0.9091                          & 50                              & 1000000                             & 0.625                          \\
Val       & 1000                               & 0.0455                          & 300                             & 300000                              & 0.1875                         \\
Test      & 1000                               & 0.0455                          & 300                             & 300000                              & 0.1875                         \\ 
\hline 
Train \#3 & 40000                              & 0.9524                          & 25                              & 1000000                             & 0.625                          \\
Val       & 1000                               & 0.0238                          & 300                             & 300000                              & 0.1875                         \\
Test      & 1000                               & 0.0238                          & 300                             & 300000                              & 0.1875                         \\ 
\hline 
Train \#4 & 50000                              & 0.9615                          & 20                              & 1000000                             & 0.625                          \\
Val       & 1000                               & 0.0192                          & 300                             & 300000                              & 0.1875                         \\
Test      & 1000                               & 0.0192                          & 300                             & 300000                              & 0.1875                         \\ \hline
Train \#5 & 100000                             & 0.9804                          & 10                              & 1000000                             & 0.625                          \\
Val       & 1000                               & 0.0098                          & 300                             & 300000                              & 0.1875                         \\
Test      & 1000                               & 0.0098                          & 300                             & 300000                              & 0.1875                         \\ \hline
Train \#6 & 200000                             & 0.9901                          & 5                               & 1000000                             & 0.625                          \\
Val       & 1000                               & 0.0050                          & 300                             & 300000                              & 0.1875                         \\
Test      & 1000                               & 0.0050                          & 300                             & 300000                              & 0.1875                         
\end{tabular}
\caption{Number and proportion of simulator queries allocated for training, validation, and testing for ex ante and interim models.}
\label{table:train_val_test_split}
\end{table*}

%% file: tables/train_payoff_statistics.tex
\begin{table}[ht]
\centering
\begin{tabular}{c|c|c|c|c|c|c}
Dataset   & $r$          & $(\vec{\sigma}, r)$ Pairs & Obs/Pair & Min Payoff & Max Payoff & Variance \\ \hline
Train \#1 & $r \leq 8$ & 10k                       & 100      & 0.129200   & 3.559154   & 0.211212 \\
Train \#2 & $r \leq 8$ & 20k                       & 50       & 0.000000   & 4.251838   & 0.283314 \\
Train \#3 & $r \leq 8$ & 40k                       & 25       & 0.000000   & 5.523588   & 0.428348 \\
Train \#4 & $r \leq 8$ & 50k                       & 20       & 0.000000   & 5.707418   & 0.497819 \\
Train \#5 & $r \leq 8$ & 100k                      & 10       & 0.000000   & 8.201802   & 0.864007 \\
Train \#6 & $r \leq 8$ & 200k                      & 5        & 0.000000   & 10.030759  & 1.597740
\end{tabular}
\caption{Payoff statistics for each of the six training datasets.}
\label{table:train_payoff_stats}
\end{table}

%% file: tables/test_payoff_statistics.tex
\begin{table}[ht]
\centering
\begin{tabular}{c|c|c|c|c|c|c}
Dataset     & $r$          & $(\vec{\sigma}, r)$ Pairs & Obs/Pair & Min Payoff & Max Payoff & Variance \\ \hline
ML Test \#1 & $r \leq 8$ & 1k                        & 300      & 0.463582   & 3.101542   & 0.160352 \\
ML Test \#2 & $r \leq 8$ & 1k                        & 500      & 0.520386   & 2.817814   & 0.150261 \\
ML Test \#3 & $r \leq 8$ & 1k                        & 700      & 0.585452   & 2.870190   & 0.146182 \\
Extra Test  & $r \leq 8$ & 757.6k                    & Many     & 0.618118   & 2.903241   & 0.128006 \\
Extra Test  & $r > 8$    & 662.9k                    & Many     & 0.010591   & 1.214115   & 0.061047
\end{tabular}
\caption{Payoff statistics for the noisy ML test set(s) and the large fine-grained grid test set.}
\label{table:test_payoff_stats}
\end{table}

%% file: tables/hyperparameter_grid.tex
\begin{table}[ht]
\centering
\begin{tabular}{l|l}
\hline
Architecture               & {[}sr64x6;hlin, sr128x6;hlin, sr256x6;hlin,                    \\
                           & r64;sr32;sr16;hlin, r128;sr64;sr32;hlin, r256;sr128;sr64;hlin, \\
                           & r256;r128;r64;hr32;hlin, r64;sr32;sr16;hsr8;hlin,              \\
                           & r128;sr64;sr32;hsr16;hlin, r256;sr128;sr64;hsr32;hlin,         \\
                           & {[}r128;r128{]}x2;hlin, {[}r128;r512;r128{]}x2;hlin,           \\
                           & {[}r64;r64{]}x3;hr16;hlin, {[}r256;r256{]}x2;hlin{]}           \\ \hline
Dropout Rate               & {[}0, 0.1, 0.2, 0.3{]}                                         \\ \hline
Batch Size                 & {[}128, 256, 512{]}                                            \\ \hline
Learning Rate ($\eta$)     & {[}0.0001, 0.0003, 0.001{]}                                    \\ \hline
Regularization ($\lambda$) & {[}0.0001, 0.001, 0.01{]}                                      \\ \hline
Layer Normalization        & {[}True, False{]}                                             
\end{tabular}
\caption{Hyperparameters tuned during the validation stage.}
\label{table:hyperparameter_grid}
\end{table}

%% file: tables/model_hyperparameters.tex
\begin{table}[h]
\centering
\begin{tabular}{l|c|c|c|c|c|c|c|c}
\multicolumn{1}{c|}{Func}    & \# $(\vec{\sigma}, r)$ & Obs & Architecture               & Batch & $\eta$ & L.N. & Drop Rt & $\lambda$ \\ \hline
\multicolumn{1}{c|}{Ex Ante} & 10k                    & 100 & {[}r256;r256{]}x2;hlin     & 256   & 3e-4   & F    & 0.3     & 1e-3      \\
                             & 20k                    & 50  & {[}r256;r256{]}x2;hlin     & 128   & 1e-4   & F    & 0.3     & 1e-4      \\
                             & 40k                    & 25  & {[}r128;r128{]}x2;hlin     & 256   & 3e-4   & F    & 0.3     & 1e-2      \\
                             & 50k                    & 20  & {[}r128;r128{]}x2;hlin     & 512   & 1e-4   & F    & 0.3     & 1e-4      \\
                             & 100k                   & 10  & {[}r128;r128{]}x2;hlin     & 512   & 1e-4   & F    & 0.3     & 1e-2      \\
                             & 200k                   & 5   & {[}r256;r256{]}x2;hlin     & 256   & 1e-4   & F    & 0.3     & 1e-2      \\ \hline
\multicolumn{1}{c|}{Interim} & 10k                    & 100 & r128;sr64;sr32;hsr16;hlin  & 256   & 1e-4   & F    & 0.1     & 1e-3      \\
                             & 20k                    & 50  & r256;sr128;sr64;hsr32;hlin & 512   & 1e-4   & F    & 0.1     & 1e-2      \\
                             & 40k                    & 25  & r256;sr128;sr64;hlin       & 512   & 1e-4   & T    & 0.1     & 1e-3      \\
                             & 50k                    & 20  & {[}r64;r64{]}x3;hr16;hlin  & 256   & 1e-4   & F    & 0.1     & 1e-3      \\
                             & 100k                   & 10  & sr128x6;hlin               & 512   & 1e-4   & F    & 0.1     & 1e-4      \\
                             & 200k                   & 5   & r128;sr64;sr32;hsr16;hlin  & 512   & 1e-4   & F    & 0.1     & 1e-3     
\end{tabular}
\caption{Summary of hyperparameters for the best-performing model for each function and training set. 
Note that "Batch" refers to batch size, $\eta$ to learning rate, "L.N." to whether layer normalization was applied after each dense layer, "Drop Rt" to the dropout rate after each dense layer, and $\lambda$ to the regularization strength.}
\label{table:model_hyperparameters}
\end{table}

%% file: tables/mix_class_rleq8.tex
\begin{table}[ht]
\centering
\begin{tabular}{lccc|c|c|c}
\multicolumn{1}{c|}{Func.} & \multicolumn{1}{c|}{\# $(\vec{\sigma}, r)$} & \multicolumn{1}{c|}{Obs/Pair} & Dead Mix & D.N.C. & Rejected NE     & Confirmed NE     \\ \hline
\multicolumn{1}{c|}{EA}  & \multicolumn{1}{c|}{200k}                      & \multicolumn{1}{c|}{5}        & 0\%           & 0\%              & 0.51\%          & 99.49\%          \\
\multicolumn{1}{l|}{}         & \multicolumn{1}{c|}{100k}                      & \multicolumn{1}{c|}{10}       & 0\%           & 0\%              & 1.93\%          & 98.07\%          \\
\multicolumn{1}{l|}{}         & \multicolumn{1}{c|}{50k}                       & \multicolumn{1}{c|}{20}       & 0\%           & 0\%              & 2.84\%          & 97.16\%          \\
\multicolumn{1}{l|}{}         & \multicolumn{1}{c|}{40k}                       & \multicolumn{1}{c|}{25}       & 0\%           & 0\%              & 0.68\%          & 99.32\%          \\
\multicolumn{1}{l|}{}         & \multicolumn{1}{c|}{20k}                       & \multicolumn{1}{c|}{50}       & 0\%           & 0\%              & 1.02\%          & 98.98\%          \\
\multicolumn{1}{l|}{}         & \multicolumn{1}{c|}{10k}                       & \multicolumn{1}{c|}{100}      & 0\%           & 0\%              & 2.50\%          & 97.50\%          \\ \hline
                              & \multicolumn{1}{l}{}                           & \multicolumn{1}{l}{}          & \textbf{0\%}  & \textbf{0\%}     & \textbf{1.58\%} & \textbf{98.42\%} \\ \hline
\multicolumn{1}{c|}{I, 1k}  & \multicolumn{1}{c|}{200k}                      & \multicolumn{1}{c|}{5}        & 0\%           & 0.85\%           & 2.39\%          & 96.76\%          \\
\multicolumn{1}{l|}{}         & \multicolumn{1}{c|}{100k}                      & \multicolumn{1}{c|}{10}       & 0\%           & 0.06\%           & 0.57\%          & 99.38\%          \\
\multicolumn{1}{l|}{}         & \multicolumn{1}{c|}{50k}                       & \multicolumn{1}{c|}{20}       & 0\%           & 0.34\%           & 2.50\%          & 97.16\%          \\
\multicolumn{1}{l|}{}         & \multicolumn{1}{c|}{40k}                       & \multicolumn{1}{c|}{25}       & 0\%           & 0\%              & 0.51\%          & 99.49\%          \\
\multicolumn{1}{l|}{}         & \multicolumn{1}{c|}{20k}                       & \multicolumn{1}{c|}{50}       & 0\%           & 0.57\%           & 1.88\%          & 97.56\%          \\
\multicolumn{1}{l|}{}         & \multicolumn{1}{c|}{10k}                       & \multicolumn{1}{c|}{100}      & 0\%           & 0.34\%           & 1.31\%          & 98.35\%          \\ \hline
                              & \multicolumn{1}{l}{}                           & \multicolumn{1}{l}{}          & \textbf{0\%}  & \textbf{0.36\%}  & \textbf{1.52\%} & \textbf{98.12\%} \\ \hline 
\multicolumn{1}{c|}{I, 2k}  & \multicolumn{1}{c|}{200k}                      & \multicolumn{1}{c|}{5}        & 0\%           & 0.06\%           & 2.56\%          & 97.39\%          \\
\multicolumn{1}{l|}{}         & \multicolumn{1}{c|}{100k}                      & \multicolumn{1}{c|}{10}       & 0\%           & 0\%              & 0.51\%          & 99.49\%          \\
\multicolumn{1}{l|}{}         & \multicolumn{1}{c|}{50k}                       & \multicolumn{1}{c|}{20}       & 0\%           & 0\%              & 2.44\%          & 97.56\%          \\
\multicolumn{1}{l|}{}         & \multicolumn{1}{c|}{40k}                       & \multicolumn{1}{c|}{25}       & 0\%           & 0\%              & 0.51\%          & 99.49\%          \\
\multicolumn{1}{l|}{}         & \multicolumn{1}{c|}{20k}                       & \multicolumn{1}{c|}{50}       & 0\%           & 0.11\%           & 1.88\%          & 98.01\%          \\
\multicolumn{1}{l|}{}         & \multicolumn{1}{c|}{10k}                       & \multicolumn{1}{c|}{100}      & 0\%           & 0\%              & 1.19\%          & 98.81\%          \\ \hline
                              & \multicolumn{1}{l}{}                           & \multicolumn{1}{l}{}          & \textbf{0\%}  & \textbf{0.03\%}  & \textbf{1.52\%} & \textbf{98.46\%}
\end{tabular}
\caption{Mixture classification results for game instances $\Gamma(\reserve)$ where $0.05 \leq r \leq 8$ in increments of 0.05.
For each learned model (row), we classify each of the 11 mixtures returned by RD for all 160 game instances and report the percentage, out of the 1760 mixed strategies, classified as dead mixture, non-convergent (Did Not Converge), rejected candidate BNE, and confirmed BNE.
We report interim results for 1,000 and 2,000 Monte Carlo marginalization samples.}
\label{table:mix_class_rleq8}
\end{table}

%% file: tables/mix_class_rgt8.tex
\begin{table}[ht]
\centering
\begin{tabular}{lccc|c|c|c}
\multicolumn{1}{c|}{Func.} & \multicolumn{1}{c|}{\# $(\vec{\sigma}, r)$} & \multicolumn{1}{c|}{Obs/Pair} & Dead Mix & D.N.C. & Rejected NE     & Confirmed NE     \\ \hline
\multicolumn{1}{c|}{EA}  & \multicolumn{1}{c|}{200k}                      & \multicolumn{1}{c|}{5}        & 4.42\%          & 0\%              & 50.00\%          & 45.58\%          \\
\multicolumn{1}{l|}{}         & \multicolumn{1}{c|}{100k}                      & \multicolumn{1}{c|}{10}       & 9.29\%          & 0\%              & 42.92\%          & 47.79\%          \\
\multicolumn{1}{l|}{}         & \multicolumn{1}{c|}{50k}                       & \multicolumn{1}{c|}{20}       & 15.71\%         & 0\%              & 42.92\%          & 41.36\%          \\
\multicolumn{1}{l|}{}         & \multicolumn{1}{c|}{40k}                       & \multicolumn{1}{c|}{25}       & 10.45\%         & 0\%              & 39.87\%          & 49.68\%          \\
\multicolumn{1}{l|}{}         & \multicolumn{1}{c|}{20k}                       & \multicolumn{1}{c|}{50}       & 11.56\%         & 0\%              & 45.58\%          & 42.86\%          \\
\multicolumn{1}{l|}{}         & \multicolumn{1}{c|}{10k}                       & \multicolumn{1}{c|}{100}      & 0.71\%          & 0\%              & 40.97\%          & 58.31\%          \\ \hline
                              & \multicolumn{1}{l}{}                           & \multicolumn{1}{l}{}          & \textbf{8.69\%} & \textbf{0\%}     & \textbf{43.71\%} & \textbf{47.60\%} \\ \hline
\multicolumn{1}{c|}{I, 1k}  & \multicolumn{1}{c|}{200k}                      & \multicolumn{1}{c|}{5}        & 0\%             & 0.1\%            & 1.75\%           & 98.18\%          \\
\multicolumn{1}{l|}{}         & \multicolumn{1}{c|}{100k}                      & \multicolumn{1}{c|}{10}       & 0\%             & 0\%              & 4.29\%           & 95.71\%          \\
\multicolumn{1}{l|}{}         & \multicolumn{1}{c|}{50k}                       & \multicolumn{1}{c|}{20}       & 0\%             & 0\%              & 0\%              & 100.00\%         \\
\multicolumn{1}{l|}{}         & \multicolumn{1}{c|}{40k}                       & \multicolumn{1}{c|}{25}       & 0\%             & 0\%              & 0.78\%           & 99.22\%          \\
\multicolumn{1}{l|}{}         & \multicolumn{1}{c|}{20k}                       & \multicolumn{1}{c|}{50}       & 0\%             & 0\%              & 4.29\%           & 95.71\%          \\
\multicolumn{1}{l|}{}         & \multicolumn{1}{c|}{10k}                       & \multicolumn{1}{c|}{100}      & 0\%             & 0\%              & 0.00\%           & 100.00\%         \\ \hline
                              & \multicolumn{1}{l}{}                           & \multicolumn{1}{l}{}          & \textbf{0\%}    & \textbf{0\%}     & \textbf{1.85\%}  & \textbf{98.14\%} \\ \hline 
\multicolumn{1}{c|}{I, 2k}  & \multicolumn{1}{c|}{200k}                      & \multicolumn{1}{c|}{5}        & 0\%             & 0\%              & 1.75\%           & 98.25\%          \\
\multicolumn{1}{l|}{}         & \multicolumn{1}{c|}{100k}                      & \multicolumn{1}{c|}{10}       & 0\%             & 0\%              & 4.29\%           & 95.71\%          \\
\multicolumn{1}{l|}{}         & \multicolumn{1}{c|}{50k}                       & \multicolumn{1}{c|}{20}       & 0\%             & 0\%              & 0\%              & 100.00\%         \\
\multicolumn{1}{l|}{}         & \multicolumn{1}{c|}{40k}                       & \multicolumn{1}{c|}{25}       & 0\%             & 0\%              & 0.78\%           & 99.22\%          \\
\multicolumn{1}{l|}{}         & \multicolumn{1}{c|}{20k}                       & \multicolumn{1}{c|}{50}       & 0\%             & 0\%              & 4.29\%           & 95.71\%          \\
\multicolumn{1}{l|}{}         & \multicolumn{1}{c|}{10k}                       & \multicolumn{1}{c|}{100}      & 0\%             & 0\%              & 0.00\%           & 100.00\%         \\ \hline
                              & \multicolumn{1}{l}{}                           & \multicolumn{1}{l}{}          & \textbf{0\%}    & \textbf{0\%}     & \textbf{1.85\%}  & \textbf{98.15\%}
\end{tabular}
\caption{Mixture classification results for game instances $\Gamma(\reserve)$ where $8.05 \leq r \leq 15$ in increments of 0.05.
For each learned model (row), we classify each of the 11 mixtures returned by RD for all 140 game instances and report the percentage, out of the 1540 mixed strategies, classified as dead mixture, non-convergent (Did Not Converge), rejected candidate BNE, and confirmed BNE.
We report interim results for 1,000 and 2,000 Monte Carlo marginalization samples.}
\label{table:mix_class_rgt8}
\end{table}
